\documentclass[aps,prd,nofootinbib]{revtex4-2}
\usepackage{fontspec}
\usepackage{amsmath}
\usepackage{amsfonts}
\usepackage{amssymb}
\usepackage{makeidx}
\usepackage{graphicx,epsfig}
\usepackage{color}
\usepackage{subfigure}
\usepackage{ulem}
\usepackage{multirow}
\usepackage{epstopdf}
\usepackage{dcolumn}
\usepackage{comment}
\usepackage{caption}
\usepackage{subcaption}
\usepackage{epstopdf}
\usepackage{dcolumn}
\usepackage{slashed}
\usepackage{makecell}
\usepackage{mathrsfs}
\usepackage{url}
\usepackage{bm}
\usepackage[section]{placeins}
\usepackage[colorlinks=true, citecolor=blue, urlcolor = blue, linkcolor= red, bookmarks=true]{hyperref}
\newcommand{\beq}{\begin{equation}}
	\newcommand{\eeq}{\end{equation}}
\newcommand{\bea}{\begin{eqnarray}}
	\newcommand{\eea}{\end{eqnarray}}

\begin{document}
	\title {Relativistic corrections to double $B_c$ meson production in $e^+e^-$ annihilation}
	
	\author{ Xiao-Peng Wang $^{(a)}$}
    \author{ Yi-Jie Li $^{(a)}$}
	\author{Guang-Zhi Xu $^{(a)}$}
	\email{ xuguangzhi@lnu.edu.cn}
	\author{Kui-Yong Liu $^{(b,a)}$}
	\email{liukuiyong@lnu.edu.cn}
	\affiliation{ {\footnotesize (a)~School of Physics, Liaoning University, Shenyang 110036, China}\\
		{\footnotesize (b)~School of Physics and Electronic Technology, Liaoning Normal University, Dalian 116029, China}}

	\date{\today}
	
	\begin{abstract}
		Within the framework of nonrelativistic QCD (NRQCD) factorization, we investigate the relativistic corrections to the production of double $B_c$ mesons in $e^+e^-$ annihilation. The study covers center-of-mass energies from the production threshold up to $2m_Z$, considering both the photon and $Z^0$-boson propagated processes. We find that the relativistic corrections are significant, with the corresponding $K$ factors of approximately 0.6. The azimuthal asymmetry, angular distribution, and transverse momentum distribution are also presented.

	\end{abstract}
	
	\maketitle
	
	\section{Introduction}
	Heavy quarkonia and the $B_c$ meson, being bound states of heavy quark pairs ($c\bar{c}$, $b\bar{b}$, $c\bar{b}$, $b\bar{c}$), serve as excellent probes for testing Quantum Chromodynamics (QCD) and the Standard Model (SM). Their theoretical description is facilitated by nonrelativistic QCD (NRQCD) factorization formalism\cite{Bodwin:1994jh}, which is avilliable due to the large heavy quark mass and the small relative velocity between $Q\bar{Q}$ (or $Q\bar{Q}'$). The pair production of heavy quarkonia in $e^+e^-$ annihilation provides a crucial testing ground for NRQCD. To match experimental data, both QCD and relativistic corrections are essential\cite{Zhang:2005cha,Gong:2007db,Dong:2012xx,Huang:2022dfw,Feng:2019zmt,He:2007te,Belle:2004abn,BaBar:2005nic}. Given that $v^2 \sim \alpha_s$, these two types of corrections are expected to be comparable in magnitude. Motivated by proposed future high-luminosity $e^+e^-$ colliders, such as the CEPC\cite{CEPCStudyGroup:2018ghi}, FCC-ee\cite{FCC:2018evy}, or a Z-factory\cite{Chang:2010am}, the leading order (LO) and next-to-leading order (NLO) QCD corrections for double heavy quarkonia production at the $Z^0$ peak have been computed\cite{Chen:2013mjb,Belov:2021ftc,Berezhnoy:2021tqb,Belov:2023hpc,Liao_2023,liaoqili2,liaoqili,Wang:2025sbx}. These studies indicate promising detection prospects and reveal that the contributions from color-octet (CO) components and relativistic corrections are also significant\cite{Wang:2025sbx}.

	The $B_c$ meson stands out in the heavy quark bound states family due to its  composition of different-flavor quarks and its unique weak decay mode. These characteristics provide clearer insights into decay and production mechanisms than other quarkonia. To date, the only observed $(b\bar{c})$ states is the ground state $B_c^+(1S)$, first discovered at the Tevatron in 1998\cite{CDF:1998ihx,CDF:1998axz} and later confirmed by LHC experiments\cite{LHCb:2012ag,LHCb:2017vlu,LHCb:2017lpu,ATLAS:2015jep,LHCb:2013xlg}, and its excited state $B_c^+(2S)$, discovered approximately fifteen years later by the ATLAS\cite{ATLAS:2014lga}, CMS\cite{CMS:2019uhm}, and LHCb collaborations\cite{LHCb:2019bem}. Theoretically, $B_c$ production has been extensively studied across various collision processes. These include direct production in $e^+e^-$ annihilation (electron production)\cite{Yang:2011ps,Yang:2013vba,Zheng:2015ixa,Zheng:2017xgj,Zheng:2018fqv,Zhang:2021ypo,Yang:2022zpc} and $\gamma\gamma$ fusion (photon production)\cite{Kolodziej:1994uu,Berezhnoy:1994bb,Berezhnoy:1995ay,Baranov:1997wy,Chen:2014xka,Dorokhov:2020nvv,Chen:2020dtu,Yang:2022yxb} at ee colliders; $gg$ fusion (hadron production) at hadron colliders\cite{Chang:1994aw,Baranov:1997wy,Chang:2004bh,Chang:2005bf,Chang:2005wd,Li:2009ug,Chen:2024dkx}; $\gamma g$ fusion (lepton-hadron production) at $ep$ colliders\cite{Berezhnoy:1997er,Hu:2024qre}; production in heavy-ion collisions   Pb+Pb at the LHC\cite{Liu:2012tn,Norbeck:2013zba}; and indirect production via the decays of heavy particles like the $W$, $Z$, and Higgs bosons, as well as the top quark\cite{Chang:2007si,Wu:2008cn,Deng:2010aq, Yang:2010yg,Liao:2011kd,Liao:2012rh,Liao:2015vqa,Liao:2018nab,Liao:2019xux,Faustov:2021itb,Belov:2021toy}, etc.

	The electron-positron production of paired $B_c$ mesons has been calculated with the collinear factorisation\cite{Wei:2018xlr}.  The LO cross section\cite{Kiselev:1993iu,Liao_2023,liaoqili2,liaoqili}, NLO QCD corrections\cite{Berezhnoy:2016etd} with  NRQCD approach, and the relativistic corrections 
	to both the $s$-channel single-photon\cite{Karyasov:2016hfm} and $t$-channel double-photon exchange processes\cite{Berezhnoy:2019jjs} within the relativistic quark model (RQM) have been investigated. However, in contrast to the extensive studies on QCD corrections to $B_c$ production or the combined QCD and relativistic corrections to charmonium and bottomonium production, relativistic effects in processes involving $B_c$ mesons have received comparatively little attention\cite{Faustov:2021itb,Belov:2021toy,Ebert:2002pp, Martynenko:2005sf,Ebert:2006xq,Lee:2010ts,Trunin:2015uma,Karyasov:2016hfm,Zhu:2017lqu,Wang:2017bgv,Zhu:2017lwi,Geng:2018qrl,Yang:2019gga,Berezhnoy:2019jjs,Dorokhov:2020nvv}. Most existing studies are based on the RQM\cite{Faustov:2021itb,Belov:2021toy,Martynenko:2005sf,Ebert:2002pp,Ebert:2006xq,Karyasov:2016hfm,Trunin:2015uma,Berezhnoy:2019jjs,Dorokhov:2020nvv}, which incorporates relativistic corrections in the amplitude for the heavy quarks' relative motion and in the bound-state wave function. 
Given that the binding effects are significantly enhanced in pair meson production,
we investigate the relativistic corrections for double $B_c$ production in $e^+e^-$ annihilation via both $\gamma^*$ and $Z^0$ propagators within the NRQCD framework. The specific processes considered are: $B_c^{*+}+B_{c}^{*-}, B_c^{*+}+B_{c}^{-}, B_c^{+}+B_{c}^{-}, B_c^{*+}+h_{bc}^{-}, B_c^{*+}+\chi_{bc0}^{-}, B_c^{*+}+\chi_{bc1}^{-}, B_c^{*+}+\chi_{bc2}^{-}, B_c^{+}+h_{bc}^{-}, B_c^{+}+\chi_{bc0}^{-}, B_c^{+}+\chi_{bc1}^{-}, B_c^{+}+\chi_{bc2}^{-}$\footnote{Processes related by charge conjugation, such as $B_c^{*+}+B_c^{-}$ and $B_c^{+}+B_c^{*-}$, have identical cross sections.}.

	The remaining of the paper is organised as follows. In Sec.~\ref{solution}, we describe the theoretical framework for the NLO($v^2$) NRQCD calculations. In Sec.~\ref{input}, we give the input parameters used in this work. In Sec.~\ref{results}, we show the results of the cross sections, the differential cross sections and the azimuthal asymmetry, etc, both with and without relativistic corrections.  And a brief conclusion is given in Sec.~\ref{conclusion}.

	\section{Method of NRQCD and relativistic corrections
	} \label{solution}
	
	\begin{widetext}	
		\begin{figure}
			\begin{tabular}{c c }
				\includegraphics[width=0.5\textwidth]{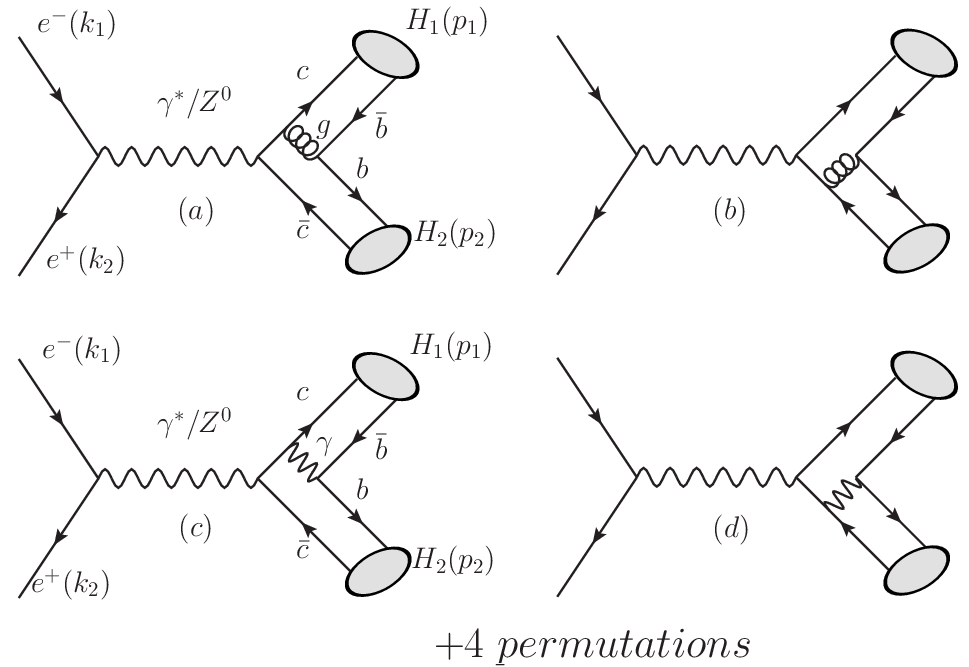}
			\end{tabular}
			\caption{Feynman diagrams for $e^-(k_1)+e^+(k_2)\rightarrow \gamma^*/Z^0\rightarrow H_1(p_1)+H_2(p_2)$ at the tree level. 
$H_1, H_2$ are $B_c^+(c\bar{b})$ and $B_c^-(b\bar{c})$ (or their respective excited states, $B_c^{*\pm}, h_{bc}^{\pm}, \chi_{bcJ}^{\pm} (J=0,1,2)$), respectively. 
The permutation diagrams can be obtained by reversing the quark lines and exchanging $b,c$ quarks. 
 (a) and (b) diagrams belong to QCD channels, (c) and (d) diagrams belong to EW channels. 
 }
			\label{feynmandia}
		\end{figure}
		
	\end{widetext}
	\FloatBarrier

	Within the NRQCD framework and up-to $\mathcal{O}(v^2)$, the production cross sections are factorized as follows\cite{Bodwin:1994jh},
	\bea
	\label{eq:factorization} \hat{\sigma}(e^++e^-{\rightarrow}H_1+H_2)&=&\sum_{n_1,n_2}(\frac{F(n_1,n_2)}{m^{d_{\mathcal{O}(n_1)-4}}m^{d_{\mathcal{O}(n_2)-4}}}\langle\mathcal{O}^{H_1}(n_1)\rangle\langle\mathcal{O}^{H_2}(n_2)\rangle\cr
	&+&\frac{G_1(n_1,n_2)}{m^{d_{\mathcal{P}(n_1)-4}}m^{d_{\mathcal{O}(n_2)-4}}}\langle\mathcal{P}^{H_1}(n_1)\rangle\langle\mathcal{O}^{H_2}(n_2)\rangle+\frac{G_2(n_1,n_2)}{m^{d_{\mathcal{O}(n_1)-4}}m^{d_{\mathcal{P}(n_2)-4}}}\langle\mathcal{O}^{H_1}(n_1)\rangle\langle\mathcal{P}^{H_2}(n_2)\rangle) 
	\label{dcs1}
	\eea
	where $n_1,n_2$ are the Fock states denoted in spectroscopic notation ${}^{2S+1}L_J^{[a]}$, with total spin $S$, orbital angular momentum $L$, total angular momentum $J$, and color configurations a=1,8 corresponding to color-singlet (CS) and CO states, respectively. $m$ denotes the heavy quark effective mass.
	$\mathcal{O}^H(n)$ and $\mathcal{P}^H(n)$ are four-fermion operators of mass dimensions $d_\mathcal{O}$ and $d_\mathcal{P}$ describing the nonperturbative  transition $n\rightarrow H$ at LO and $\mathcal{O}(v^2)$. $\langle\mathcal{O}^H(n)\rangle$ and $\langle\mathcal{P}^H(n)\rangle$ are the respective long-distance matrix elements (LDMEs), and $F(n_1,n_2)$ and $G_i(n_1,n_2)$ are the appropriate short-distance-coefficients (SDCs).  The SDCs describe the short-distance production of two Fock states $(c\bar{b})[n_1]$ and $(b\bar{c})[n_2]$ and the LDMEs describe the hadronization of the Fock state $n_{1,2}$ into the physical $B_c$ mesons $H_{1,2}$.
	The SDCs are perturbatively calculable and the LDMEs can be obtained by the potential model\cite{Eichten:1994gt,Liao:2014rca,Eichten:2019gig} and potential NRQCD\cite{Brambilla:1999xf} etc.
	
    In the subsequent discussion, we will see that for double $B_c$ production, the contribution of the CO channels are negligible compared with the CS channels. Within the CS framework,
	the relevant four-quark operators for the production of $H=B_c, B_c^*,\ldots$ are defined as\cite{Bodwin:1994jh}
	\bea
	\mathcal{O}^{H}(^1S_0^{[1]})&=&\chi^{\dagger} \psi(a^{\dagger}_Ha_H)\psi^{\dagger} \chi,\cr
	\mathcal{P}^{H}(^1S_0^{[1]})&=&\frac{1}{2}[\chi^{\dagger}\ \psi(a^{\dagger}_Ha_H)\psi^{\dagger} (-\frac{i}{2}\overleftrightarrow{\boldsymbol{D}})^2\chi+h.c],\cr
	\mathcal{O}^{H}(^3S_1^{[1]})&=&\chi^{\dagger}\boldsymbol{\sigma}^i\psi(a^{\dagger}_Ha_H)\psi^{\dagger}\boldsymbol{\sigma}^i\chi,\cr
	\mathcal{P}^{H}(^3S_1^{[1]})&=&\frac{1}{2}[\chi^{\dagger}\boldsymbol{\sigma}^i\psi(a^{\dagger}_Ha_H)\psi^{\dagger}\boldsymbol{\sigma}^i(-\frac{i}{2}\overleftrightarrow{\boldsymbol{D}})^2\chi+h.c.],\cr
	&\cdots&
	\eea
	where $\psi(\chi)$ is the Pauli spinor that annihilates (creates) a heavy quark (antiquark), $\boldsymbol{\sigma}^i(i=1,2,3)$ are the Pauli matrices, and $\overleftrightarrow{\boldsymbol{D}}=\overleftarrow{\boldsymbol{D}}-\overrightarrow{\boldsymbol{D}}$, with $\overrightarrow{\boldsymbol{D}}$ being the space component of the covariant derivative $D^\mu$.

	For $S$ and $P$ wave states, the LDMEs at LO are related to the values of radial functions and the first derivatives of radial functions at the origin,
	\bea
	\frac{\langle\mathcal{O}^{B_c^*}(^3S_1^{[1]})\rangle}{2N_c\times3}=\frac{|R_S(0)|^2}{4\pi},~	\frac{\langle\mathcal{O}^{B_c}(^1S_0^{[1]})\rangle}{2N_c}=\frac{|R_S(0)|^2}{4\pi},\cr
	\frac{\langle\mathcal{O}^{\chi_{bcJ}}(^3P_J^{[1]})\rangle}{2N_c\times(2J+1)}=\frac{3|R'_P(0)|^2}{4\pi},~	\frac{\langle\mathcal{O}^{h_{bc}}(^1P_1^{[1]})\rangle}{2N_c\times3}=\frac{3|R'_P(0)|^2}{4\pi}
	\eea
	It's convenient to define matrix elements $\langle v^2\rangle$ instead of the LDMEs at $\mathcal{O}(v^2)$\cite{Bodwin:2002cfe,Bodwin:2007fz,Guo:2011tz}
	\beq
	\langle v^2\rangle\equiv\frac{\langle\mathcal{P}^H(n)\rangle}{m^2\langle\mathcal{O}^H(n)\rangle}
	\eeq

	The SDCs can be obtained via the matching condition between perturbative QCD and NRQCD calculations,
	\bea
	\hat{\sigma}(e^++e^-{\rightarrow}(c\bar{b})_1+(b\bar{c})_2)|_{\text{pert ~QCD}}&=& [\frac{F(n_1,n_2)}{m^{d_{\mathcal{O}(n_1)-4}}m^{d_{\mathcal{O}(n_2)-4}}}\langle0|\mathcal{O}^{(c\bar{b})_1}(n_1)|0\rangle\langle0|\mathcal{O}^{(b\bar{c})_2}(n_2)|0\rangle\cr
	&+&\frac{G_1(n_1,n_2)}{m^{d_{\mathcal{P}(n_1)-4}}m^{d_{\mathcal{O}(n_2)-4}}}\langle0|\mathcal{P}^{(c\bar{b})_1}(n_1)|0\rangle\langle0|\mathcal{O}^{(b\bar{c})_2}(n_2)|0\rangle\cr&+&\frac{G_2(n_1,n_2)}{m^{d_{\mathcal{O}(n_1)-4}}m^{d_{\mathcal{P}(n_2)-4}}}\langle0|\mathcal{O}^{(c\bar{b})_1}(n_1)|0\rangle\langle0|\mathcal{P}^{(b\bar{c})_2}(n_2)|0\rangle]|_{\text{pert~NRQCD}}
	\label{matching}
	\eea
	The left-hand side of Eq. (\ref{matching}) can be computed easily by using the covariant projection method\cite{Bodwin:2002cfe}.

	The Feynman diagrams for the exclusive $B_c$ pair production $e^-e^+ \to \gamma^*/Z^0 \to H_1^+(p_1) + H_2^-(p_2)$ are displayed in Fig. 1. 
    Here, we only present diagrams for s-channel processes. The t-channel double photon fragmentation mechanism dominates the production of $J/\psi$ and $\Upsilon$ pairs \cite{Bodwin:2002kk,Bhatnagar:2024ykb,Wang:2025sbx}, as the photon can fragment into a ${}^3S_1$ quarkonium state. By contrast, similar to the case of $\eta_{c/b}$ pair production where the photon cannot fragment into a ${}^1S_0$ state due to C-parity conservation, that of the $B_c$ pair production proceeds only via the non-fragmentation mechanism, since a photon cannot fragment into a $c\bar{b}$ quark pair. Subsequent calculations indicate that the t-channel contributions are negligible.
    In the diagrams, $H_1$ and $H_2$ are the positively and negatively charged $B_c$ mesons, with the quark contents of $c(p_c)\bar{b}(p_{\bar{b}})$ and $b(p_b)\bar{c}(p_{\bar{c}})$, respectively.
	The momenta of the heavy quarks in the mesons are defined as follows,
	\bea
	p_c&=&\frac{E_c}{E_c+E_{\bar{b}}}p_1+q_1,~ p_{\bar{b}}=\frac{E_{\bar{b}}}{E_c+E_{\bar{b}}}p_1-q_1,\cr
	p_b&=&\frac{E_b}{E_{\bar{c}}+E_b}p_2+q_2,~ p_{\bar{c}}=\frac{E_{\bar{c}}}{E_{\bar{c}}+E_b}p_2-q_2
	\eea 
	$~E_{c/\bar{b}}=\sqrt{m_{c/b}^2+\vec{q}_1^2}$ and $E_{b/\bar{c}}=\sqrt{m_{b/c}^2+\vec{q}_2^2}$ are the energies of constituent quarks in the meson's rest frame, where $\vec{q}_i(i=1,2)$ denote the three-momenta of the quarks, and each can be expressed as the product of the reduced mass $\mu_{H}=\frac{m_cm_b}{m_c+m_b}$ and the effective quark velocity $\vec{v}$. Meanwhile, $q_{i}$ are the corresponding four-momenta, subject to constraints $q_i\cdot p_i=0$. 
	\bea
	|\vec{q}_i| = 2\mu_{H}v_i, 
	\eea 
	
	The full QCD scattering amplitude can be expressed as 	
	\bea
	A(q_1,q_2)&=&\sum_{ k_il_i}\langle s_{c};s_{\bar{b}}|S_1S_{1z} \rangle\langle3,k_1;\bar{3},l_1|1\rangle \langle s_{\bar{c}};s_{b}|S_2S_{2z} \rangle\langle3,k_2;\bar{3},l_2|1\rangle\cr&\times&\mathcal{A}(e^++e^-\rightarrow c_{k_1}+\bar{b}_{l_1}+b_{k_2}+\bar{c}_{l_2})
	\label{amp1}
	\eea
	with $\mathcal{A}(e^+e^-\rightarrow c+\bar{b}+b+\bar{c})$ being the standard Feynman amplitude and $\langle3,k_i;\bar{3},l_i\rangle=\delta_{k_il_i}/\sqrt{N_c}$ being the Clebsch-Gordan coefficient in SU(3) color space for the $(bc)$ pair to be projected onto the CS state applying the covariant projector method\cite{He:2007te,He:2009uf,Guo:2011tz,Xu:2012am,Li:2013csa,Xu:2014zra, Wang:2025sbx,Zhu:2017lqu,Wang:2017bgv,Zhu:2017lwi,Yang:2019gga}.
	The projector operator for $H_1$ in spin-triplet state is given as follows,
	\bea
	\label{proj}
	\mathbb{P}_1&=& \sum_{s_{c},s_{\bar{b}}} \langle s_{c};s_{\bar{b}}|SS_z \rangle v(p_{\bar{b}};s_{\bar{b}})\bar{u}(p_{c};s_c)\cr
	&=&\frac{1}{2\sqrt{2}\sqrt{E_c+m_c}\sqrt{E_{\bar{b}}+m_{b}}}(-\slashed{p}_{\bar{b}}+m_{b})\slashed{\epsilon}_1 \frac{\slashed{p}_{c}+\slashed{p}_{\bar{b}}+E_c+E_{\bar{b}}}{E_c+E_{\bar{b}}}(\slashed{p}_{c}+m_c)\otimes(\frac{\bm{1}_c}{\sqrt{N_c}})
    \eea
	where Dirac spinors are normalized as $\bar{u}u=2m_c,-\bar{v}v=2m_b$. $\epsilon_{1}$ is unit spin polarization vector. 
For spin-singlet state,  $\epsilon_{1}$ is replaced by $\gamma_5$. $\bm{1}_c$ represents the unit color matrix, $N_c$ is the number of colors in QCD. The projector operator for $H_2$ can be  obtained  by making the  replacement  $\bar{b}\rightarrow \bar{c},c\rightarrow b$, and $m_c\leftrightarrow m_b$  in Eq.~(\ref{proj}).
	The complete color factor for QCD diagrams (a, b) is $\frac{N_c^2-1}{2N_c}=C_F$,  pure EW diagrams (c, d) is 1, and that for t-channel double photon exchange processes which will be discussed in Sec. \ref{dcs}, is 1.

	To expand Eq. (\ref{amp1}) as a double series in $q_1$ and $q_2$, it is convenient to define
	
		\bea
	A_{\alpha_1\cdots\alpha_m,\beta_1\cdots\beta_n}=\sqrt{\frac{m_1}{E_1}}\sqrt{\frac{m_2}{E_2}}\big[\frac{\partial^{m+n}A(q_1,q_2)}{\partial q_1^{\alpha_1}\cdots\partial q_1^{\alpha_m}\partial q_2^{\beta_1}\cdots\partial q_2^{\beta_n}}\Big|_{q_1=q_2=0}\big]
	\eea 
	where the factor $\sqrt{m_i/E_i}$ compensates the relativistic normalization
\footnote{These normalization factors will be $\sqrt{2m_1(E_c+E_{\bar{b}})}$ and $\sqrt{2m_2(E_b+E_{\bar{c}})}$ when Dirac spinors are normalized as $u^\dagger u=v^\dagger v=1$, and the projection operator in Eq.~(\ref{proj}) is divided by a factor of $2\sqrt{E_cE_{\bar{b}}}$\cite{Bodwin:2002cfe,Zhu:2017lqu,Wang:2017bgv,Zhu:2017lwi,Yang:2019gga}.}
of the $(bc)_i$ pair with $m_1=m_2=2\mu_H,
	E_1=2E_cE_{\bar{b}}/(E_c+E_{\bar{b}}), E_2=2E_bE_{\bar{c}}/(E_b+E_{\bar{c}})$.	Such that
	\bea
	A(q_1,q_2)&=&A_{0,0}+q_1^{\alpha_1}A_{\alpha_1,0} +q_2^{\beta_1}A_{0,\beta_1} \cr
	&+&\frac{1}{2}q_1^{\alpha_1}q_1^{\alpha_2}A_{\alpha_1\alpha_2,0} +\frac{1}{2}q_2^{\beta_2}q_2^{\beta_2}A_{0,\beta_1\beta_2} +\cdots.
	\eea
	Here, the subscript ``0" means no derivatives of $q_i$ to the amplitude.
	
	For S-wave states, only even powers of $q_i$ contribute in the expansion. The replacement below is adopted in our calculations.
	\bea
	q_i^\mu q_i^\nu\rightarrow\frac{|\vec{q}_i|^2}{3}\Pi_i^{\mu\nu}
	\label{tensor}
	\eea
	with $\Pi_i^{\mu\nu}=-g^{\mu\nu}+\frac{p_i^\mu p_i^\nu}{p_i^2}$. 
	For P-wave states, only odd powers of $q_i$ contribute. The corresponding replacements are given as follows.
	\bea
	q_i^\mu&\rightarrow&|\vec{q}_i|\epsilon_{L_z}^\mu(p_i)\cr
	q_i^{\mu}q_i^{\nu}q_i^{\xi}&\rightarrow&\frac{|\vec{q}_i|^3}{5}\Pi^{\mu\nu\xi}_i=\frac{|\vec{q}_i|^3}{5}(\Pi_i^{\mu\nu}\epsilon_{L_z}^{\xi}+\Pi_i^{\nu\xi}\epsilon_{L_z}^{\mu}+\Pi_i^{\xi\mu}\epsilon_{L_z}^{\nu})
	\eea
	where $\epsilon_{L_z}(p_i)$ denotes the orbital polarisation vector of the P-wave meson.
	
	As for the kinematics,  
	the Mandelstam variables are defined as follows,
	\bea
	s=(k_1+k_2)^2, t=(k_1-p_1)^2, u=(k_1-p_2)^2 
	\eea
	$t, u$ depend on $|\vec{q}_{i}|^2$. We define $t_0,u_0$ as the values at $|\vec{q}_{i}|^2=0$ (non-relativistic result), therefore $t, u$ are expanded up-to $\mathcal{O}(v^2)$ as follows.
	
	\bea
	t=t_0+\frac{t_0+M^2}{4M^2-s}\frac{M^2}{m_cm_b}(|\vec{q}_{1}|^2+|\vec{q}_{2}|^2) 
	+ \mathcal{O}(v^4)\\
	u=u_0+\frac{u_0+M^2}{4M^2-s}\frac{M^2}{m_cm_b}(|\vec{q}_{1}|^2+|\vec{q}_{2}|^2) 
	+ \mathcal{O}(v^4),
	\eea	
	where $M=m_c+m_b$.
The master formulas for the SDCs in the matching condition of Eq. (\ref{matching}) are given by,
	\bea	 
	\frac{F(n_1,n_2)}{m^4}&=&\frac{1}{8s}\int \sum|\mathcal{M}|^2d\phi_2^{(0)}\\
	\frac{G_1(n_1,n_2)}{m^6}&=&\frac{1}{8s}\int (f_{ps}*\sum|\mathcal{M}|^2+\sum|N_1|^2)d\phi_2^{(0)}\\
	\frac{G_2(n_1,n_2)}{m^6}&=&\frac{1}{8s}\int (f_{ps}*\sum|\mathcal{M}|^2+\sum|N_2|^2)d\phi_2^{(0)}
	\eea
where $|\mathcal{M}|^2$,$|N_i|^2$ denote the amplitudes squared at $\mathcal{O}(v^0)$ and $\mathcal{O}(v^2)$, respectively. $1/(8s)$ is the flux factor $1/(2s)$ multiplied by the spin-average factor $1/4$ for the initial electron-positron pair. $d\phi_2^{(0)}$ is the two-body phase space integral element at $\mathcal{O}(v^0)$, and 
$f_{ps}=-\frac{r}{m_cm_b(1-4r)}$ with $r=(m_c+m_b)^2/s$ comes from the relativistic expansion to the phase space factor
\footnote{The phase space integral element is expanded as $d\phi_2=\frac{|\vec{p}_{1}|}{32\pi^2s^{3/2}}\frac{p_f}{p_i}d\Omega = d\phi_2^{(0)}\big[1+f_{ps}*(|\vec{q}_1|^2+|\vec{q}_2|^2)+\mathcal{O}(v^4)\big]$.}.
The amplitudes squared are explicitly expressed as below,
 	\bea
 |\mathcal{M}|^2&=&
 \begin{cases}
 	|A_{0,0}|^2\Big|_{\vec{q}_1^2=\vec{q}_2^2=0}\\
 	\epsilon^{*\beta_1'}_{L_z}\epsilon^{\beta_1}_{L_z}A^*_{0,\beta_1'}A^{}_{0,\beta_1}\Big|_{\vec{q}_1^2=\vec{q}_2^2=0}
 	\label{msqu0}\\
 \end{cases}\\
 |N_1|^2&=&
 \begin{cases}
 	[\frac{\partial}{\partial \vec{q}_1^2}( |A_{0,0}|^2)+\frac{1}{3}\Pi_1^{\alpha_1\alpha_2}\text{Re}[A^*_{0,0}A_{\alpha_1\alpha_2,0}]]\Big|_{\vec{q}_1^2=\vec{q}_2^2=0}  \\
 	[\frac{\partial}{\partial \vec{q}_1^2}( \epsilon^{*\beta_1'}_{L_z}\epsilon^{\beta_1}_{L_z}A^*_{0,\beta_1'}A^{}_{0,\beta_1})+\frac{1}{3}\Pi_1^{\alpha_1\alpha_2}\epsilon^{*\beta_1'}_{L_z}\epsilon^{\beta_1}_{L_z}\text{Re}[A^*_{0,\beta_1'}A^{~}_{\alpha_1\alpha_2,\beta_1}]]\Big|_{\vec{q}_1^2=\vec{q}_2^2=0}  \label{msqu1} \\
 \end{cases}\\
 |N_2|^2&=&
 \begin{cases}
 	[\frac{\partial}{\partial \vec{q}_2^2}( |A_{0,0}|^2)+\frac{1}{3}\Pi_2^{\beta_1\beta_2}\text{Re}[A^*_{0,0}A_{0,\beta_1\beta_2}]]\Big|_{\vec{q}_1^2=\vec{q}_2^2=0}  \\
 	[\frac{\partial}{\partial \vec{q}_2^2}( \epsilon^{*\beta_1'}_{L_z}\epsilon^{\beta_1}_{L_z}A^*_{0,\beta_1'}A^{~}_{0,\beta_1})+\frac{1}{15}\epsilon^{*\beta_1'}_{L_z} \Pi_{2}^{\beta_1\beta_2\beta_3}\text{Re}[A^*_{0,\beta_1'}A^{~}_{0,\beta_1\beta_2\beta_3}]]\Big|_{\vec{q}_1^2=\vec{q}_2^2=0}
 \end{cases}
 \label{msqu2}
 \eea
where the first and second rows of Eqs. (\ref{msqu0},\ref{msqu1},\ref{msqu2}) correspond to the production of double S-wave $B_c$ mesons and ``S+P''-wave $B_c$ mesons, respectively.

	\section{Input parameters}
	\label{input}
	 
    We use the following input parameters in the numerical calculations\cite{ParticleDataGroup:2024cfk,Berezhnoy:2021wrc}  
	\bea
	\alpha=1/137, 	m_c&=&1.5~GeV, m_b=4.8~GeV, m_Z=91.1876~GeV,\cr
	~\Gamma_Z&=&2.4952~GeV,
	~\sin^2\theta_w=0.2312,~v^2=0.15 
	\eea
	$ m_c, m_b$ are the masses of charm  and bottom quark, $m_Z,~\Gamma_Z$ are the mass and width of $Z^0$ boson, $\theta_w$ is the Weinberg angle. The values of the square of the effective velocity $v^2$ are estimated as\cite{Berezhnoy:2021wrc}:
	\bea
	v_{i}^2=\frac{\langle T\rangle}{2\mu_H}, \textrm{for}\ i=1,2.
	\eea
where $\langle T\rangle$ is the averaged kinematic energy of quark inside the $B_c$ meson and is estimated to $0.35~GeV$ adopted from Ref. \cite{Gershtein:1994jw}.

The values of the original wave functions could be found in several Refs. \cite{Ebert:2011jc,Liao:2014rca,Eichten:2019gig,Berezhnoy:2019jjs,Berezhnoy:2021wrc}. Here, we use the set from Ref. \cite{Eichten:2019gig},
	\bea
	|R_S(0)|^2=1.994~GeV^3,~|R'_P(0)|^2=0.3083~GeV^5.
	\eea
	The matrix elements $\langle v^2\rangle$ adhere to the velocity power scaling rules and are of the order of $v^2$ as follows,
	\bea
	v^2=\langle v^2\rangle[1+\mathcal{O}(v^4)].
	\eea
	
	The running of strong coupling constant is taken at one-loop accuracy, 
	\bea
	\alpha_s(\mu)=\frac{2\pi}{(11-2/3n_f)\ln(\mu/\Lambda_{QCD})}
	\eea
	with the active quark flavors $n_f=3$, $\Lambda_{QCD}=251~MeV$ and $\alpha_s(m_Z)=0.1184$.
	
	\section{Results and discussion}
	\label{results}
	
	\subsection{Cross section and differential cross section}
	\label{dcs}
\begin{widetext}
		\begin{figure*}[htbp]
			\begin{tabular}{c c c}
				\includegraphics[width=0.333\textwidth]{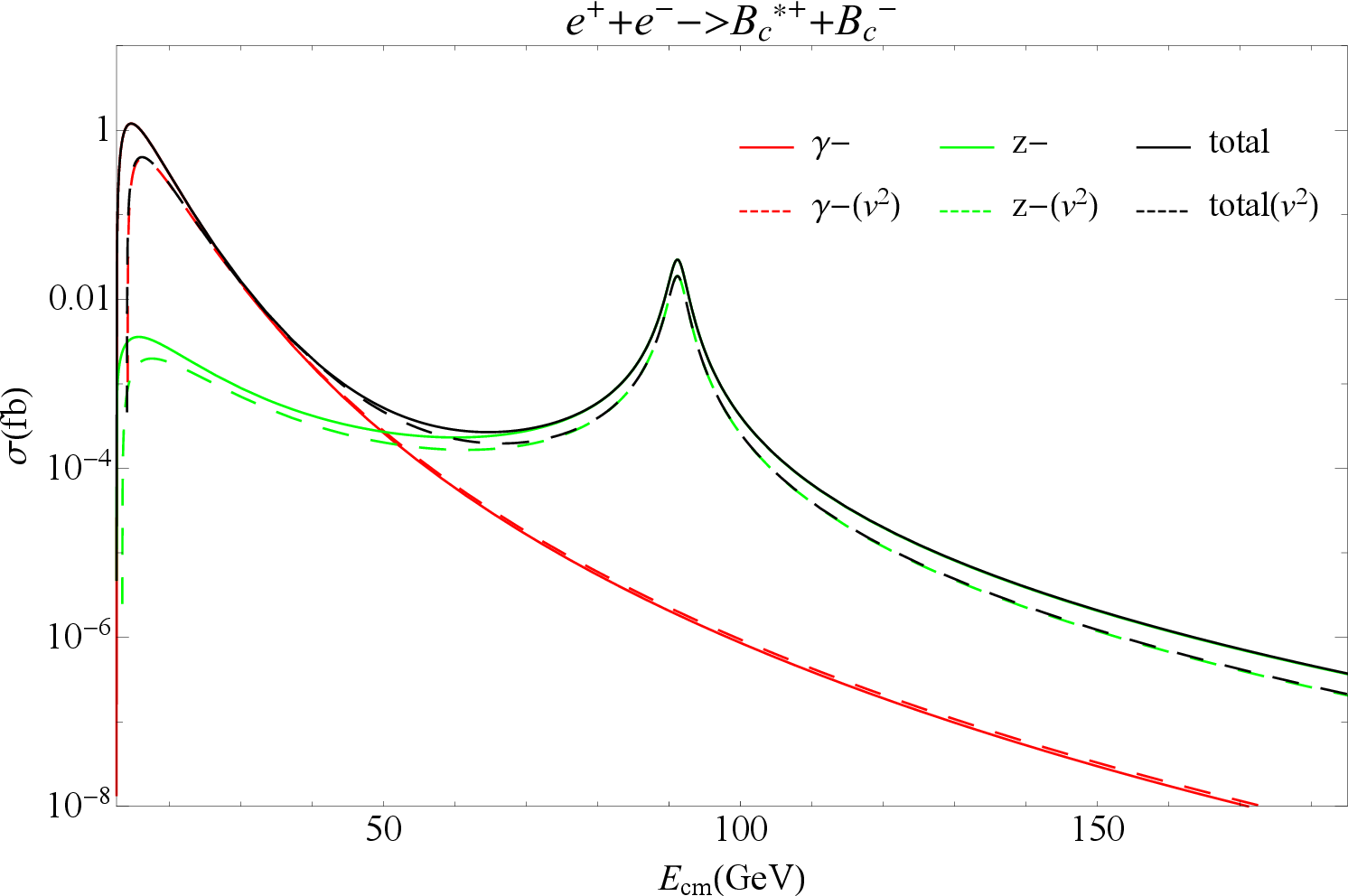}
				\includegraphics[width=0.333\textwidth]{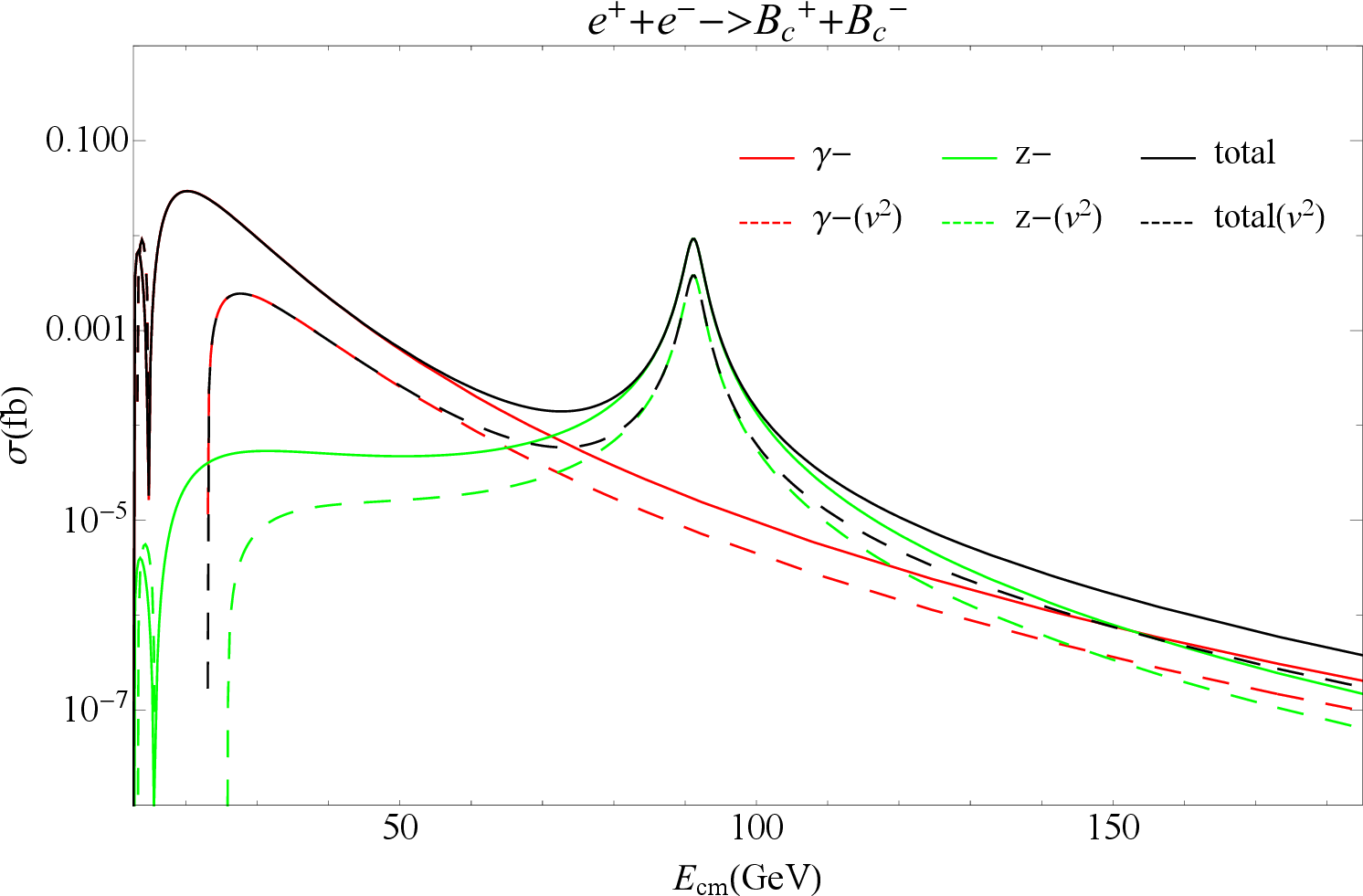}
				\includegraphics[width=0.333\textwidth]{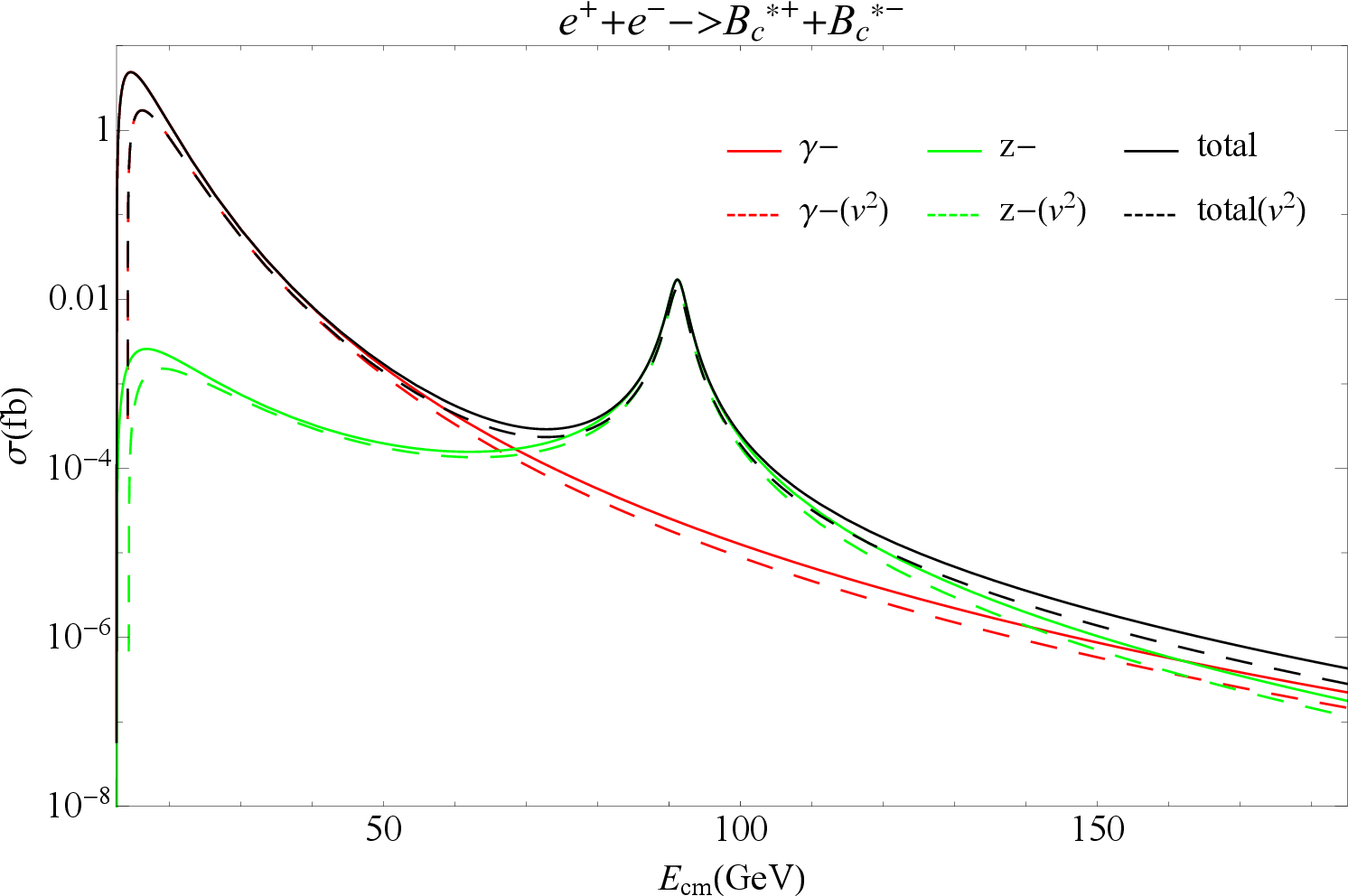}
			\end{tabular}
			\begin{tabular}{c c c}
				\includegraphics[width=0.333\textwidth]{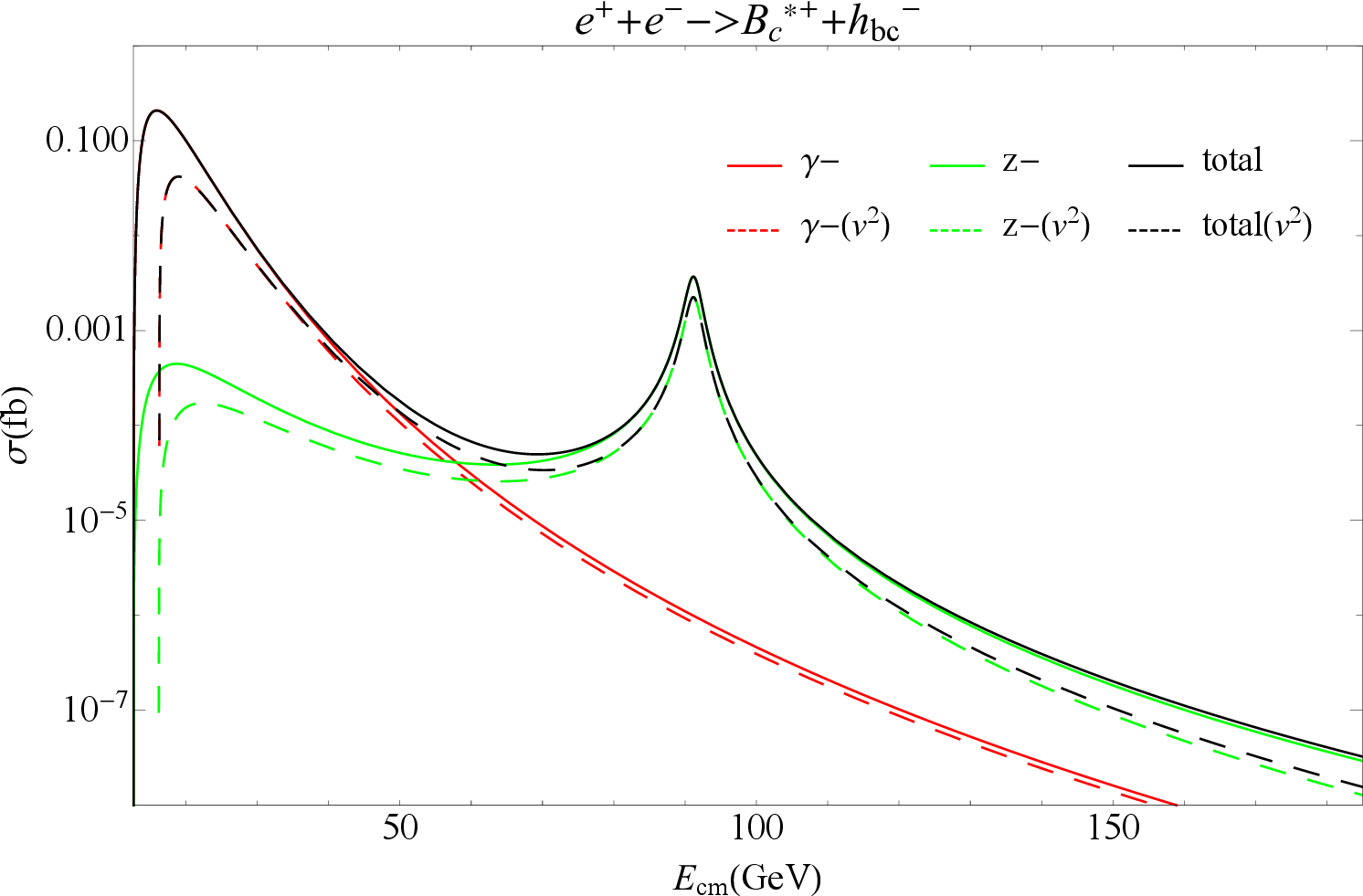}
				\includegraphics[width=0.333\textwidth]{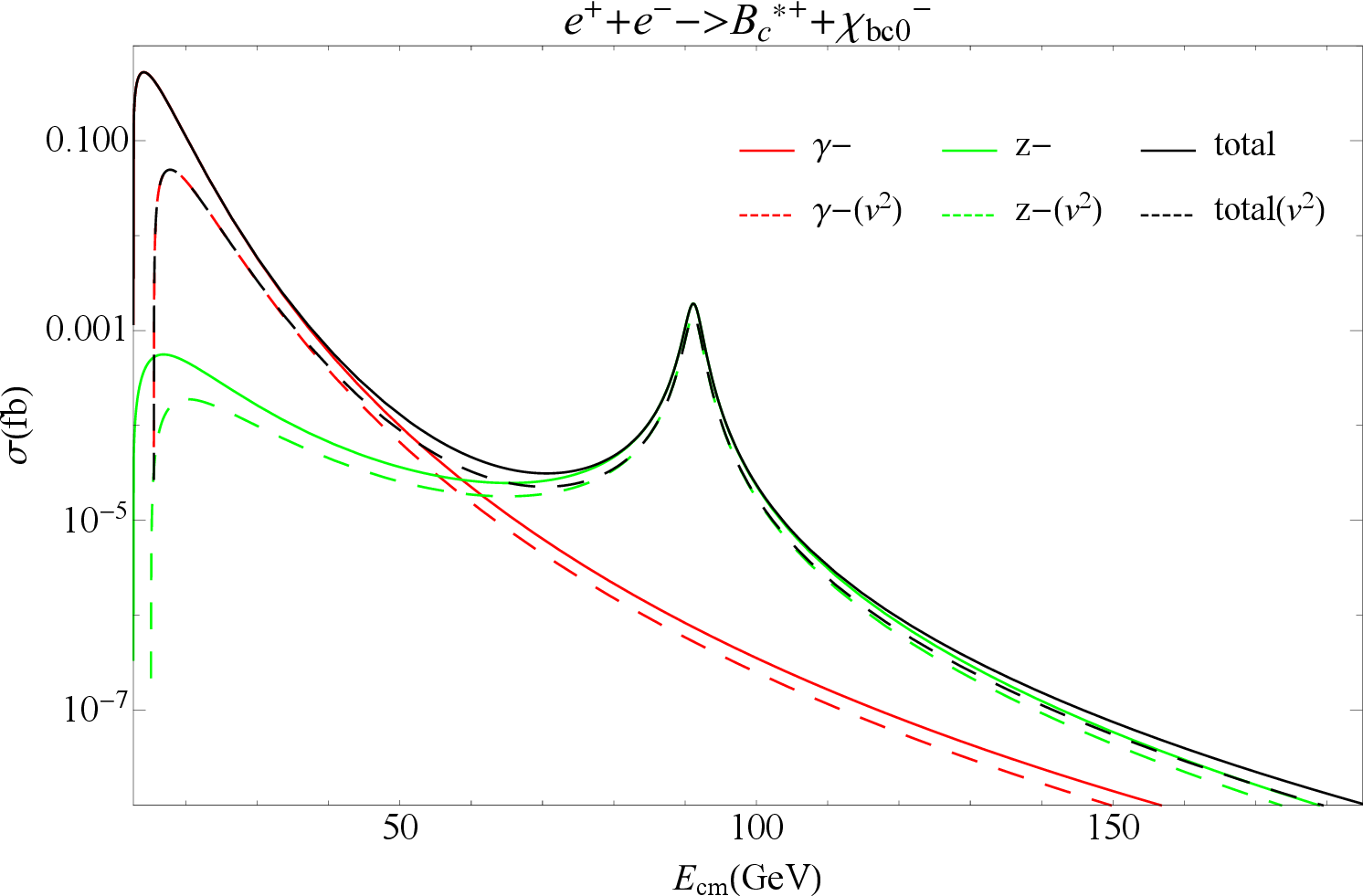}
				\includegraphics[width=0.333\textwidth]{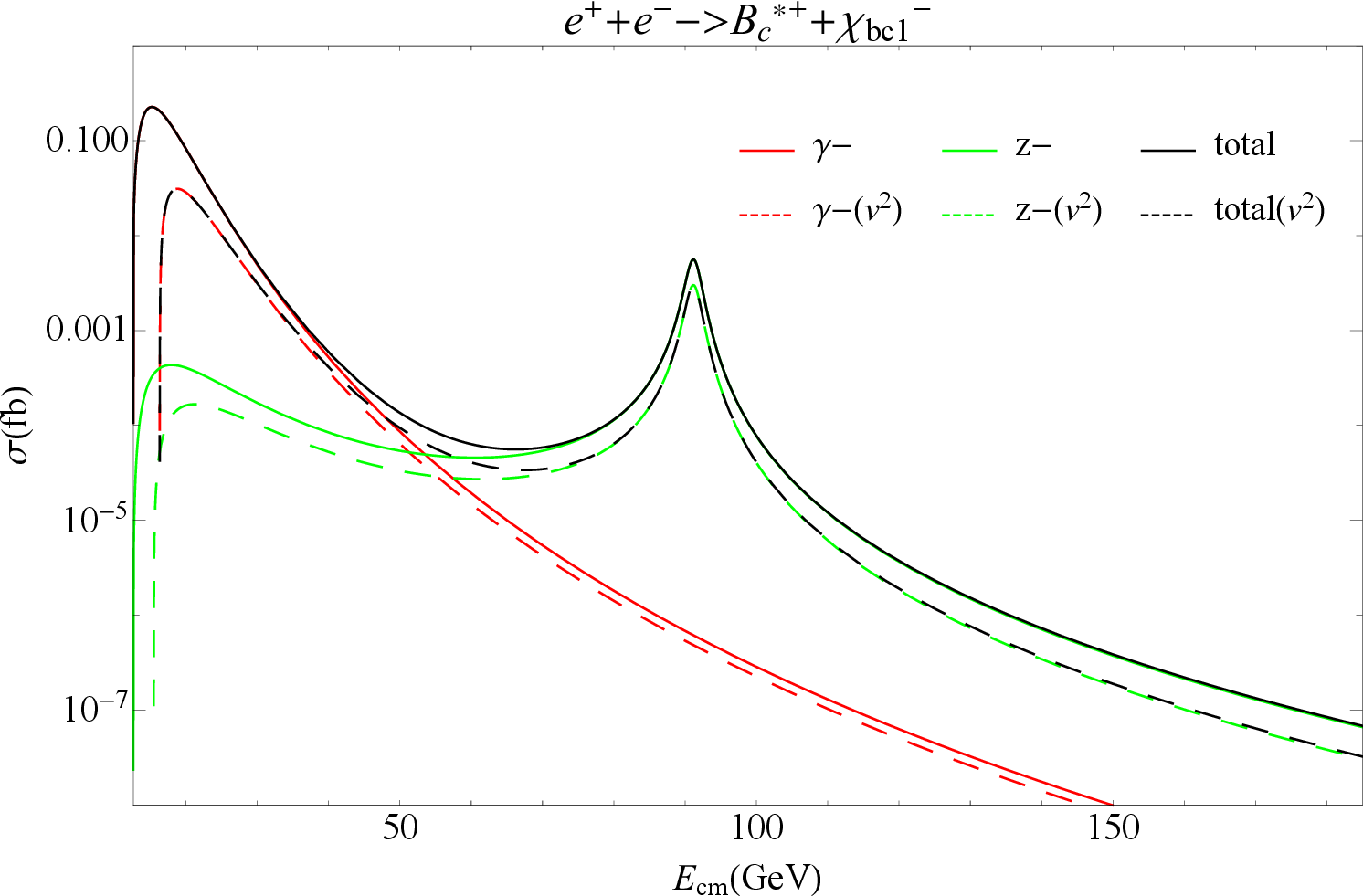}
			\end{tabular}
			\begin{tabular}{c c c}
				\includegraphics[width=0.333\textwidth]{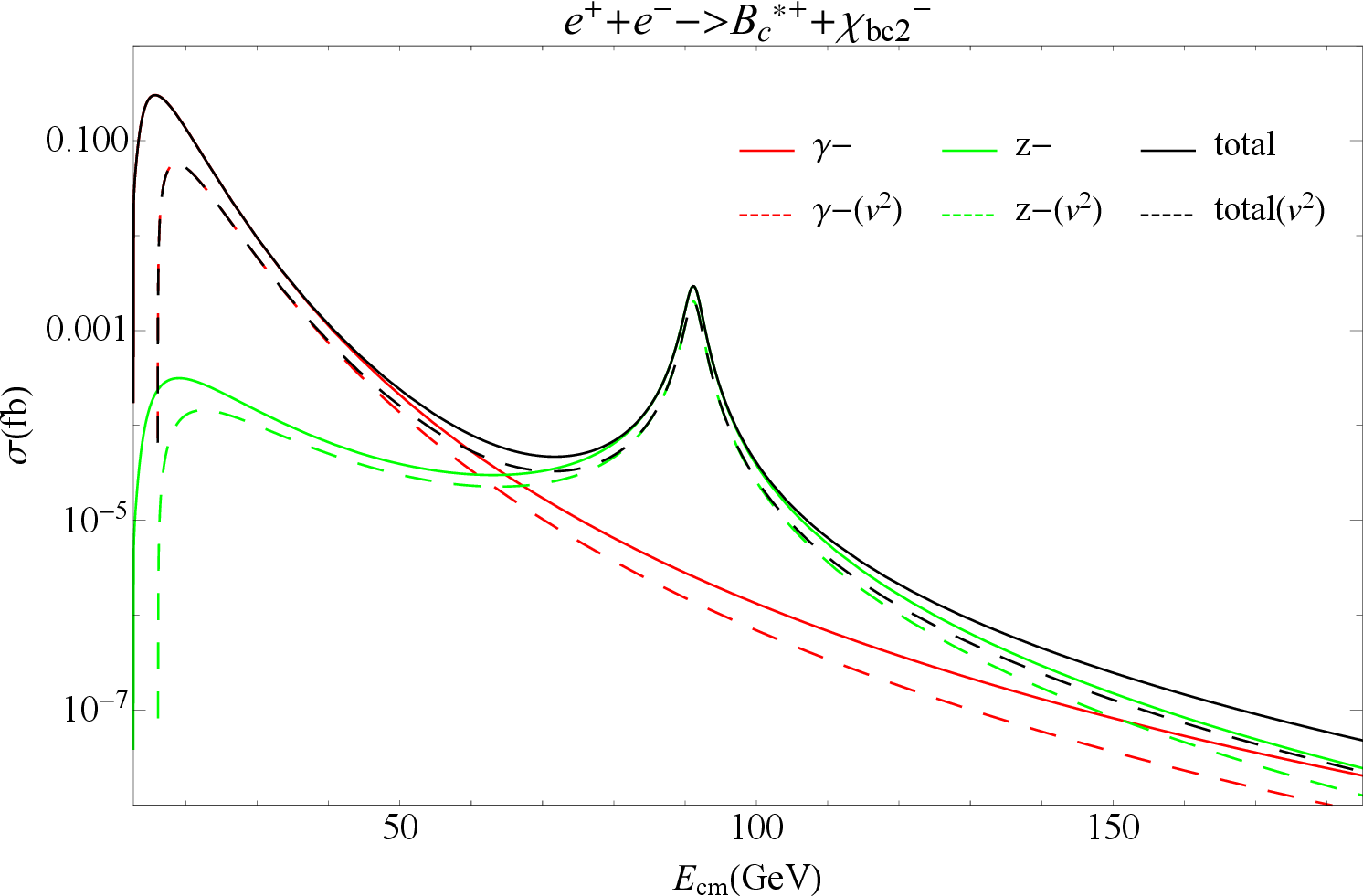}
				\includegraphics[width=0.333\textwidth]{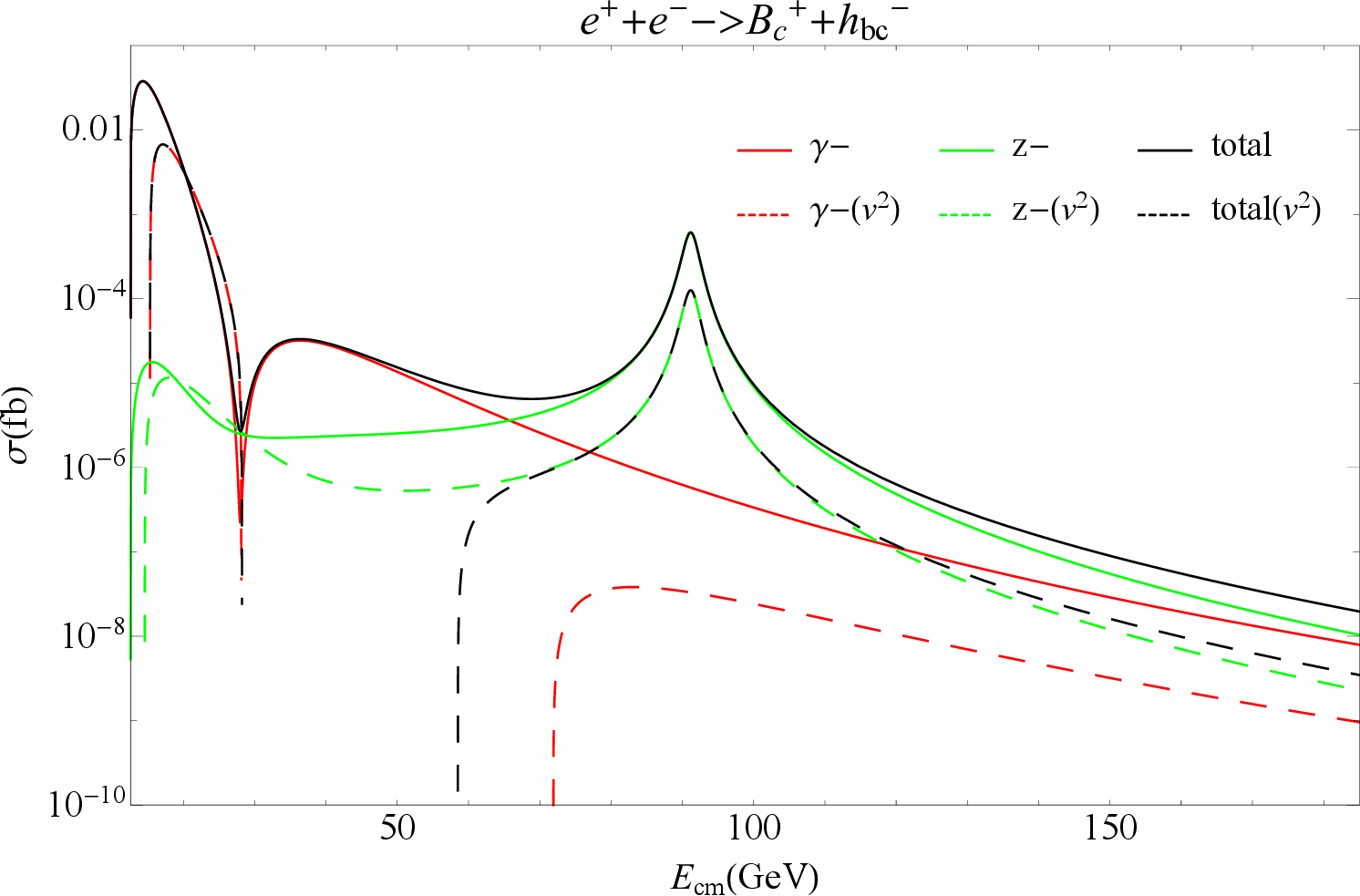}
				\includegraphics[width=0.333\textwidth]{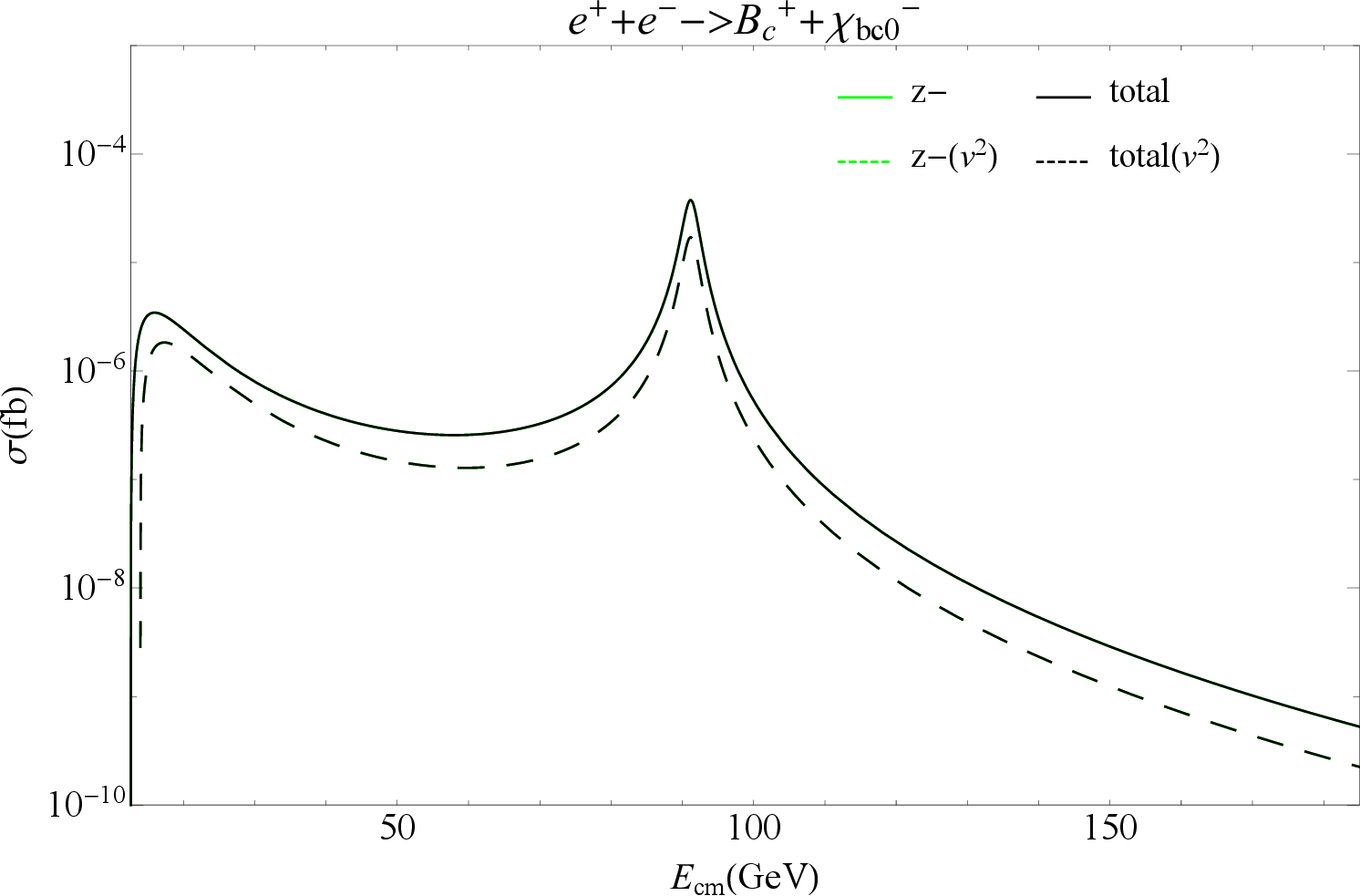}
			\end{tabular}
			\begin{tabular}{c c c}
				\includegraphics[width=0.333\textwidth]{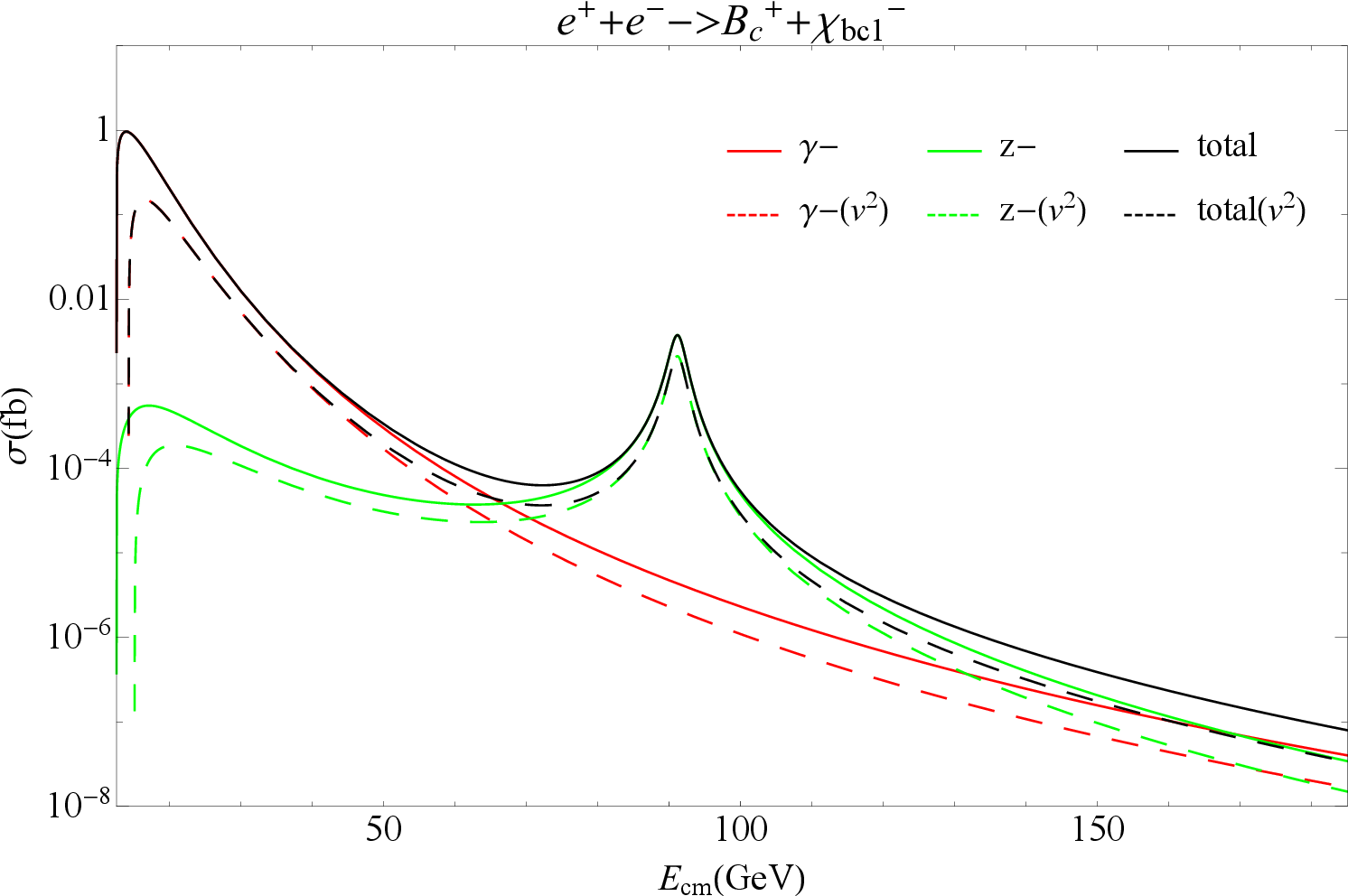}
				\includegraphics[width=0.333\textwidth]{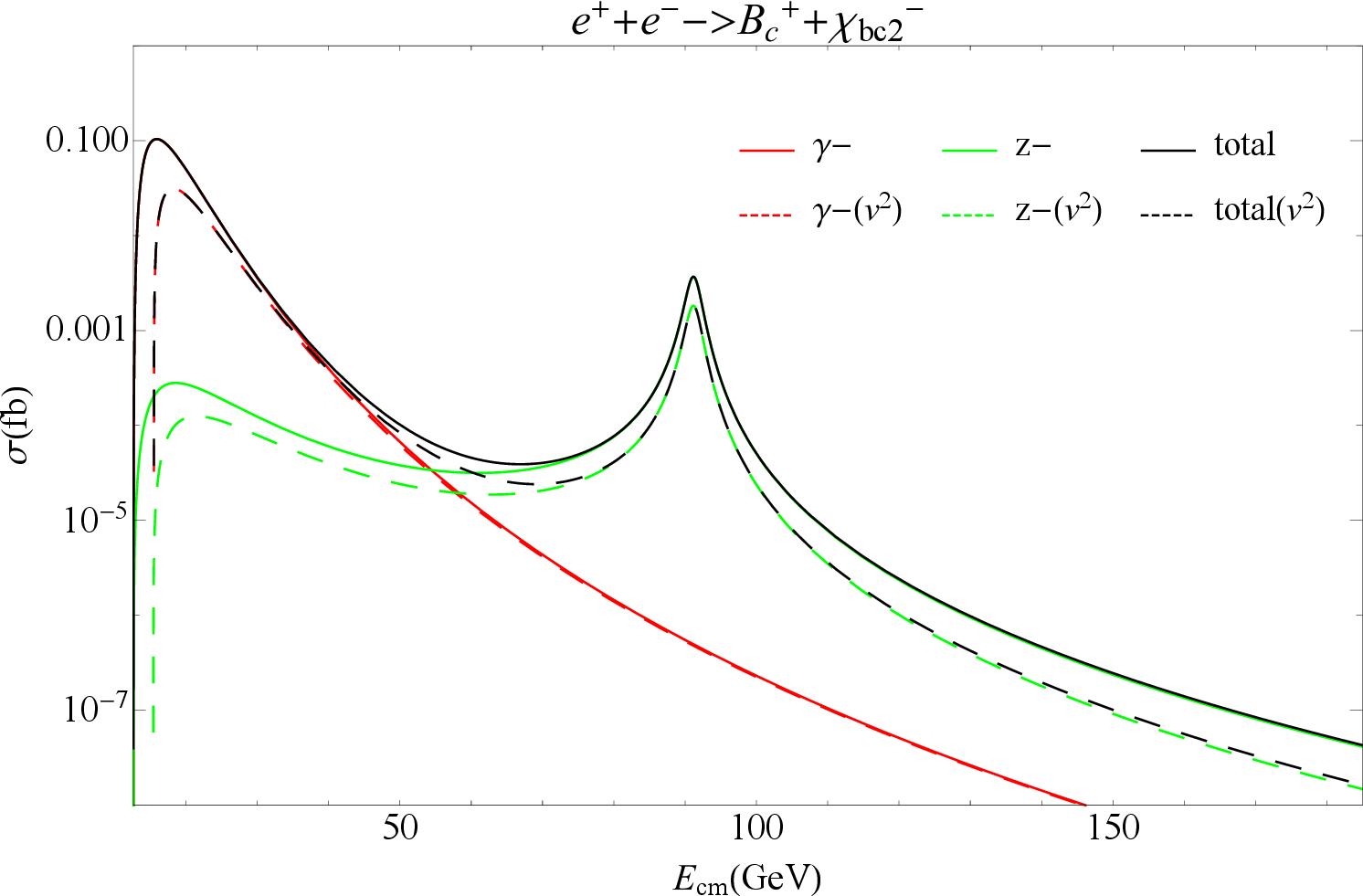}
			\end{tabular}
			\caption{(Color online) Cross section $\sigma$ versus c.m. energy  $\sqrt{s}~(E_{cm}=\sqrt{s})$. The red, green, black color lines correspond to $\gamma$-propagated processes,   $Z$-propagated processes, and the sum of both, respectively. The solid line represents LO results, while the dashed line represents NLO($v^2$) results.   }
			\label{cs}
		\end{figure*}
	\end{widetext}

	Fig. \ref{feynmandia} shows four types of processes, denoted as $\gamma$-$\gamma$, $\gamma$-$g$, $Z$-$\gamma$, and $Z$-$g$. These labels specify the nature of the two propagators in each process.
	The contributions from the pure electroweak (EW) diagrams ($\gamma^*$/ $Z^0$-$\gamma$ type) are several orders of magnitude smaller than those from the QCD ($\gamma^*$/ $Z^0$-$g$ type) diagrams. 
This suppression factors can be estimated by $[Q_cQ_b\alpha/(C_F\alpha_s)]^2\sim 10^{-4}$, which arise from the coupling constants, the charge of the heavy quark, and color factors. 
For this reason, previous studies in Refs.~\cite{Karyasov:2016hfm,Berezhnoy:2016etd,Liao_2023,liaoqili2,liaoqili} considered only the QCD processes.
	Here, we do not show the results of the four types of processes ($\gamma$-$\gamma$, $\gamma$-$g$, $Z$-$\gamma$, and $Z$-$g$), but instead classify them into two types, i.e., $\gamma$-propagated ($\gamma$-$\gamma$, $\gamma$-$g$) and $Z^0$-propagated ($Z$-$\gamma$, $Z$-$g$) processes.
	Therefore we investigate the paired $B_c$ production via $\gamma^*/Z^0$-propagators separately, as in Refs. \cite{Chen:2013mjb,Liao_2023,liaoqili}, and show the cross sections as a function of the center-of-mass (c.m.) energyin Fig.~\ref{cs}. One can see that the $\gamma$-propagated processes are dominant in the low energy region, while the $Z^0$-propagated processes are dominant around the Z-factory energy region due to the resonance effect. Moreover, the interference between different mechanisms (e.g., fragmentation and nonfragmentation \cite{Lu:2024cjp}) may sometimes cause destructive contributions.
	We also calculated the interference contributions and found they were moderate, with the ratio $  (\sigma_{\text{total}}-\sigma_{\gamma}-\sigma_{Z})/\sigma_{\text{total}}$ generally ranging from -5\% to 5\%.
	In the following sections, we will only discuss the total results (the sum of both processes, including the interference).

\begin{widetext}
		\begin{figure*}[htbp]
			\begin{tabular}{c c c}
				\includegraphics[width=0.333\textwidth]{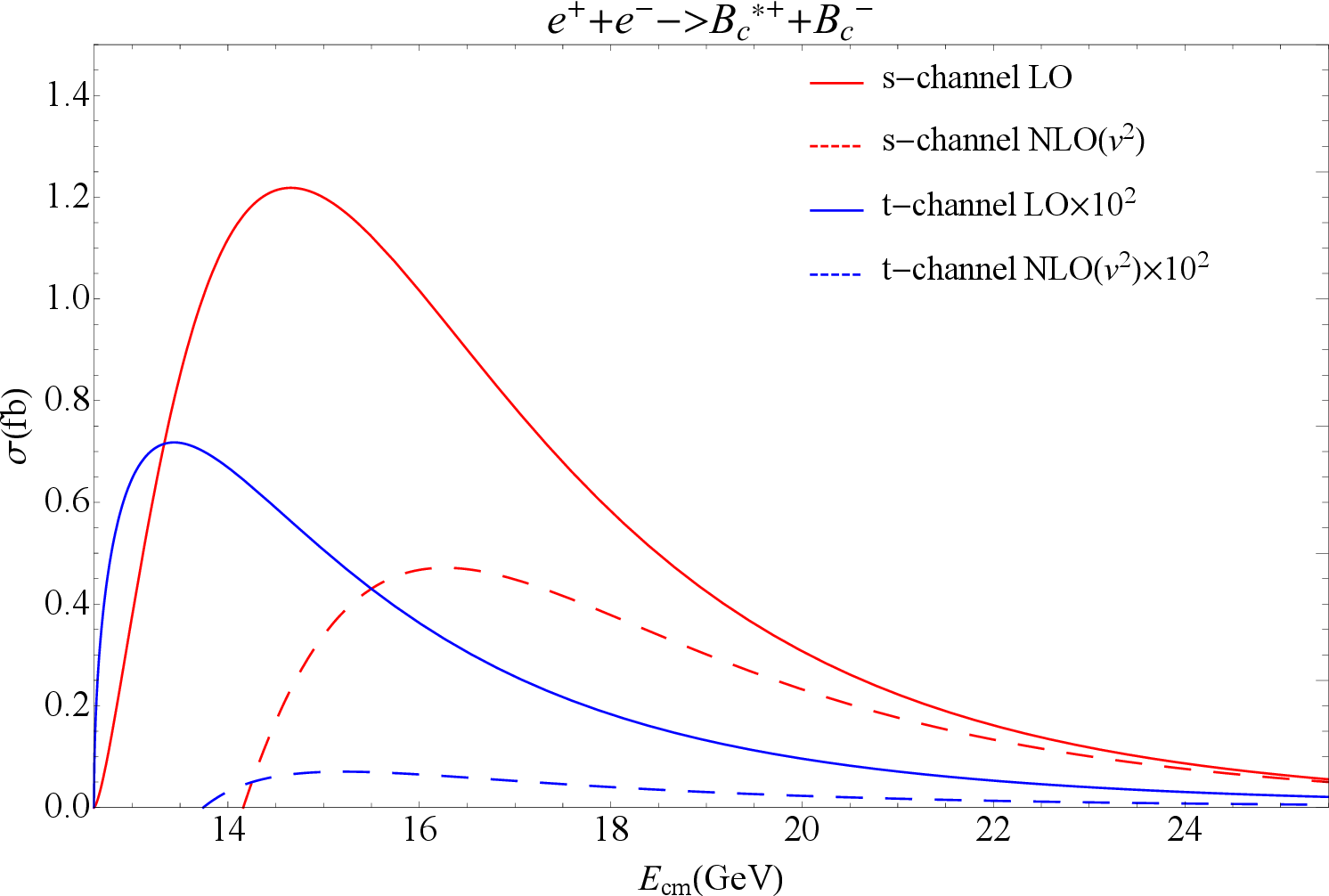}
				\includegraphics[width=0.333\textwidth]{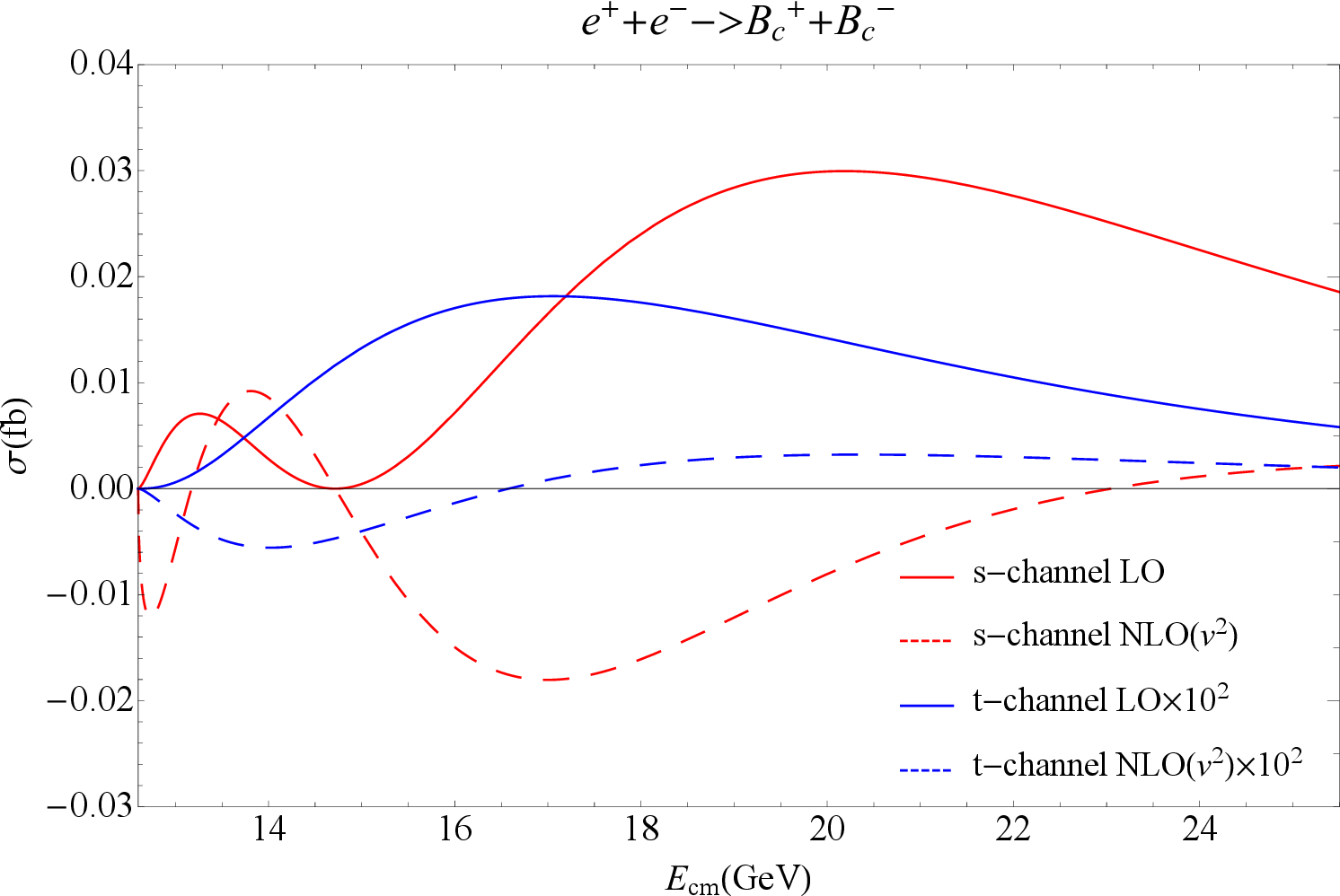}	
				\includegraphics[width=0.333\textwidth]{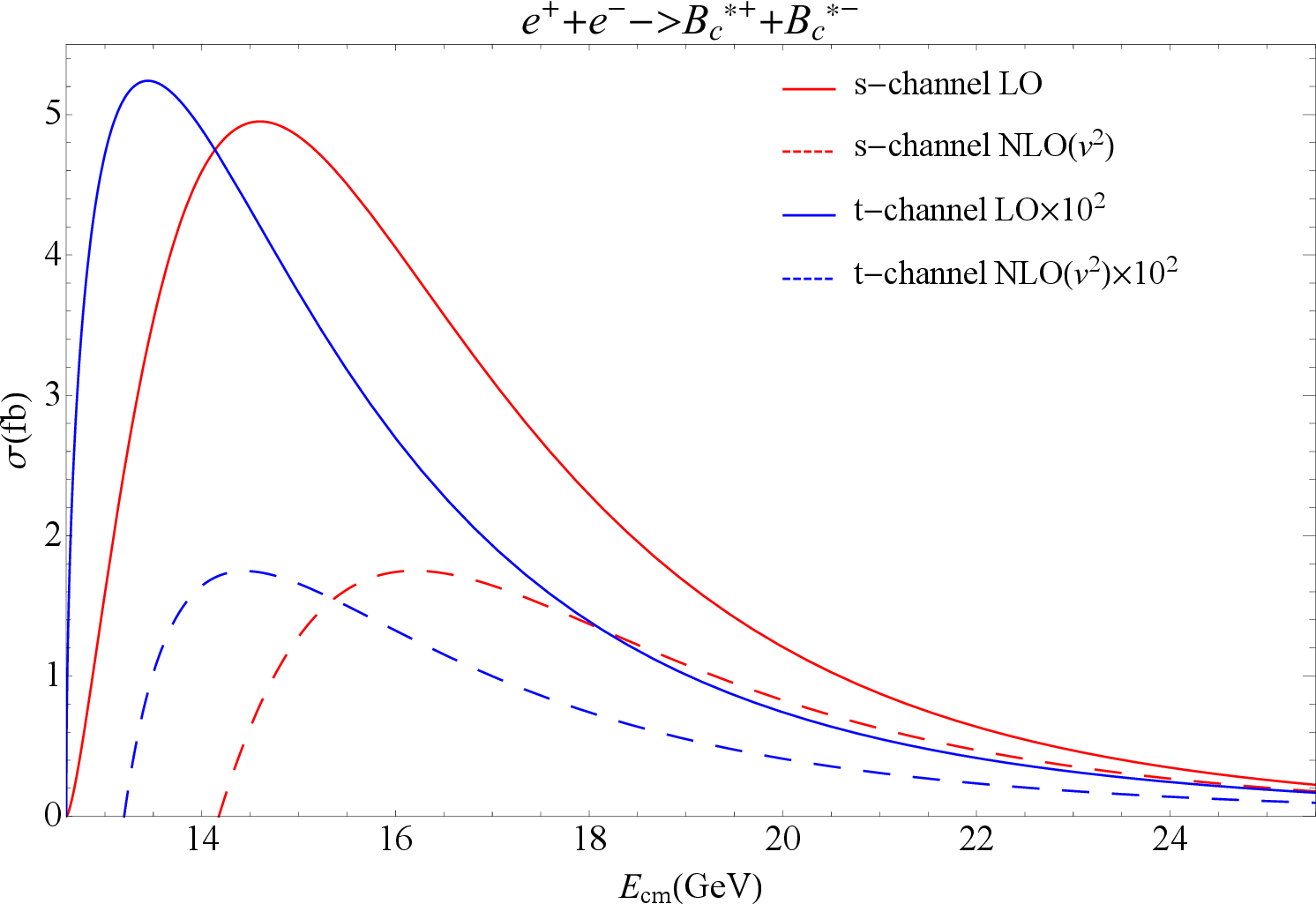}
			\end{tabular}	
			\caption{(Color online) Cross section $\sigma$ versus c.m. energy  $\sqrt{s}~(E_{cm}=\sqrt{s})$ for the s-channel $\gamma$-$g$ type process and t-channel process.   The  red and blue lines represent s-channel and t-channel processes, respectively. The solid and dashed lines represent LO and NLO($v^2$) results, respectively. The t-channel results are scaled up by a factor of 100.   }
			\label{sgag}
		\end{figure*}
	\end{widetext}

In contrast to Ref.~\cite{Karyasov:2016hfm}, which studied the production of double $S$-wave $B_c$ mesons via a virtual photon within the RQM, we present our results for the same mechanism within the NRQCD frame (the $s$-channel $\gamma$-$g$ type) in Fig.~\ref{sgag}. The NRQCD method yields results are qualitatively consistent with those in the RQM calculations, namely the relativistic corrections have negative effects. The corresponding numerical cross sections at $\sqrt{s} = 22\ \text{GeV}$ are provided in Table~\ref{s22gev}.

	\begin{table}
		\caption{Production  cross sections of $B_c^{*+}+B_c^{-}$, $B_c^{+}+B_c^{-}$, and $B_c^{*+}+B_c^{*-}$ production for s-channel $\gamma-g$ type process and t-channel double-photon process. LO and NLO($v^2$) means leading order and next-to-leading order in the $v^2$ expansions, respectively.    }
		\begin{tabular}{|c|c|c|c|c|}
			\hline
			\multicolumn{5}{|c|}{s-channel} \\
			\hline
			$\sqrt{s}=22.0 ~ GeV$&LO(this work)&NLO($v^2$)(this work) &LO\cite{Karyasov:2016hfm}~&NLO($v^2$)\cite{Karyasov:2016hfm}  \\
			\hline
			$B_c^{*+}+B_c^{-}$&  0.162 $fb$   &   0.133 $fb$   &0.10$fb$ &0.04$fb$\\
			\hline
			$B_c^{+}+B_c^{-}$&0.028$ fb$  & -0.002$fb$ & 0.10$fb$& 0.03$fb$\\
			\hline
			$B_c^{*+}+B_c^{*-}$&  0.638 $fb$  &  0.472 $ fb$ &2.14$fb$ &0.58$fb$\\
			\hline
			\hline
			\multicolumn{5}{|c|}{t-channel} \\
			\hline
			$\sqrt{s}=22.0 ~ GeV$&LO(this work)&NLO($v^2$)(this work) &LO\cite{Berezhnoy:2019jjs}~&NLO($v^2$)\cite{Berezhnoy:2019jjs}  \\
			\hline
			$B_c^{*+}+B_c^{-}$& 0.53$\times10^{-3}fb$   &  0.13$\times10^{-3}fb$ & 0.18$\times10^{-3}fb$& 0.11$\times10^{-3}fb$ \\
			\hline
			$B_c^{+}+B_c^{-}$&0.10$\times10^{-3}fb$  &0.03$\times10^{-3}fb$  &0.03$\times10^{-3}fb$ &0.02$\times10^{-3}fb$\\
			\hline
			$B_c^{*+}+B_c^{*-}$&  4.16$\times10^{-3}fb$  &  2.35$\times10^{-3}fb$& 1.43$\times10^{-3}fb$& 0.46$\times10^{-3}fb$\\
			\hline
		\end{tabular}
		\label{s22gev}
	\end{table}

	The contributions of the t-channel double photon exchange have also been studied\cite{Berezhnoy:2019jjs,Negash:2025wbe}. We calculated the t-channel double photon production cross sections and found they were at least two orders of magnitude smaller than the corresponding s-channel cross sections for all $11$ processes considered herein. This is due to a suppression factor of $\alpha^2/\alpha_s^2$ ($\sim10^{-3}$). Therefore, these contributions can be safely neglected.
	Nevertheless, for comparison with Ref.~\cite{Berezhnoy:2019jjs}, which studied the t-channel double-photon production mechanism within the RQM, we also present our relativistic results for 
t-channel processes within NRQCD frame in Fig.~\ref{sgag}. 
Like the RQM results of Ref.~\cite{Berezhnoy:2019jjs}, our study demonstrates that relativistic effects lead to negative corrections.
The remaining differences may originate from distinct theoretical frameworks and input parameters (e.g., quark masses and LDMEs). The numerical results at $\sqrt{s} = 22\ \mathrm{GeV}$ are presented in Table~\ref{s22gev}.
	
	Furthermore, regarding $B_c^{+}+B_c^{-}$ production, Ref.~\cite{Berezhnoy:2016etd} proposed that higher-order contributions, such as two-loop QCD crorrections or relativistic corrections, might resolve the problem of negative cross sections observed in the NLO QCD calcualtions. However, Fig.~\ref{sgag} shows that this issue persists even when relativistic corrections are included. Additionally, the interferences between s- and t-channels were calculated and also failed to resolve the problem.

	The numerical results of cross sections for all processes are presented in Table \ref{TCS1}, in which we list the cross sections for different renormalization scales at three energy points ( $m_Z/2, m_Z, 2m_Z$), for both LO and NLO in $v^2$ expansions.
	The LO results are consistent with that in Ref. \cite{Berezhnoy:2016etd} when using the same input parameters
\footnote{The difference between the cross sections presented herein and those in Ref. \cite{Berezhnoy:2016etd} stems from the choice of LDMEs. Specifically, Ref. \cite{Berezhnoy:2016etd} adopts the LDMEs for $B_c$ mesons as $\langle\mathcal{O}\rangle_{B_c}=\frac{M}{2}f_{B_c}^2$ with $f_{B_c}^2=0.16~ GeV^2$. The factor $\langle\mathcal{O}\rangle_{B_c}/(2Nc)$ is the counterpart of $|R_S(0)|^2/(4\pi)$ adopted in our calculations. This discrepancy in the LDME-related factors leads to a cross section ratio of $\simeq0.3$.}.
For $B_c^{+}+B_c^{-}$ production via the $\gamma$-propagator, the LO cross section takes a zero at an energy point $s_0$ near the threshold or in the low energy region  
\footnote{This dip behavior also appears in $e^+e^-\rightarrow\gamma^*/Z^0\rightarrow \chi_{Q0}+\gamma$ production\cite{Chen:2013mjb,Liao:2021ifc} due to an overall suppression factor of $(s-12m_Q^2)^2$. For instance, tanking $m_Q=m_c=1.5 ~GeV$, the corresponding energy point is $\sqrt{s_0}\simeq5.2~ GeV$.} 
, and the value of $s_0$ is obtained as follows,
	\bea
	s_0(B_c^{+}+B_c^{-})&=&\frac{2M^2(1-4r_c+6r_c^2-4r_c^3+3r_c^4)}{{r_c(1-3r_c+5r_c^2-3r_c^3)}},\sqrt{s_0}\simeq14.7~ GeV
	\eea
	where $r_c=m_c/M$, $M=m_c+m_b$, which is consistent with Ref.~\cite{Berezhnoy:2016etd}. 
	In contrast, for $B_c^{+}+h_{bc}^{-}$ production (via the $\gamma$-propagator), the LO cross section at $\sqrt{s}\simeq28.0~GeV$ is not strictly zero ($\simeq1.3\times10^{-7} fb$).

\begin{widetext}
		\begin{figure*}[htbp]
			\begin{tabular}{c c c}
				\includegraphics[width=0.333\textwidth]{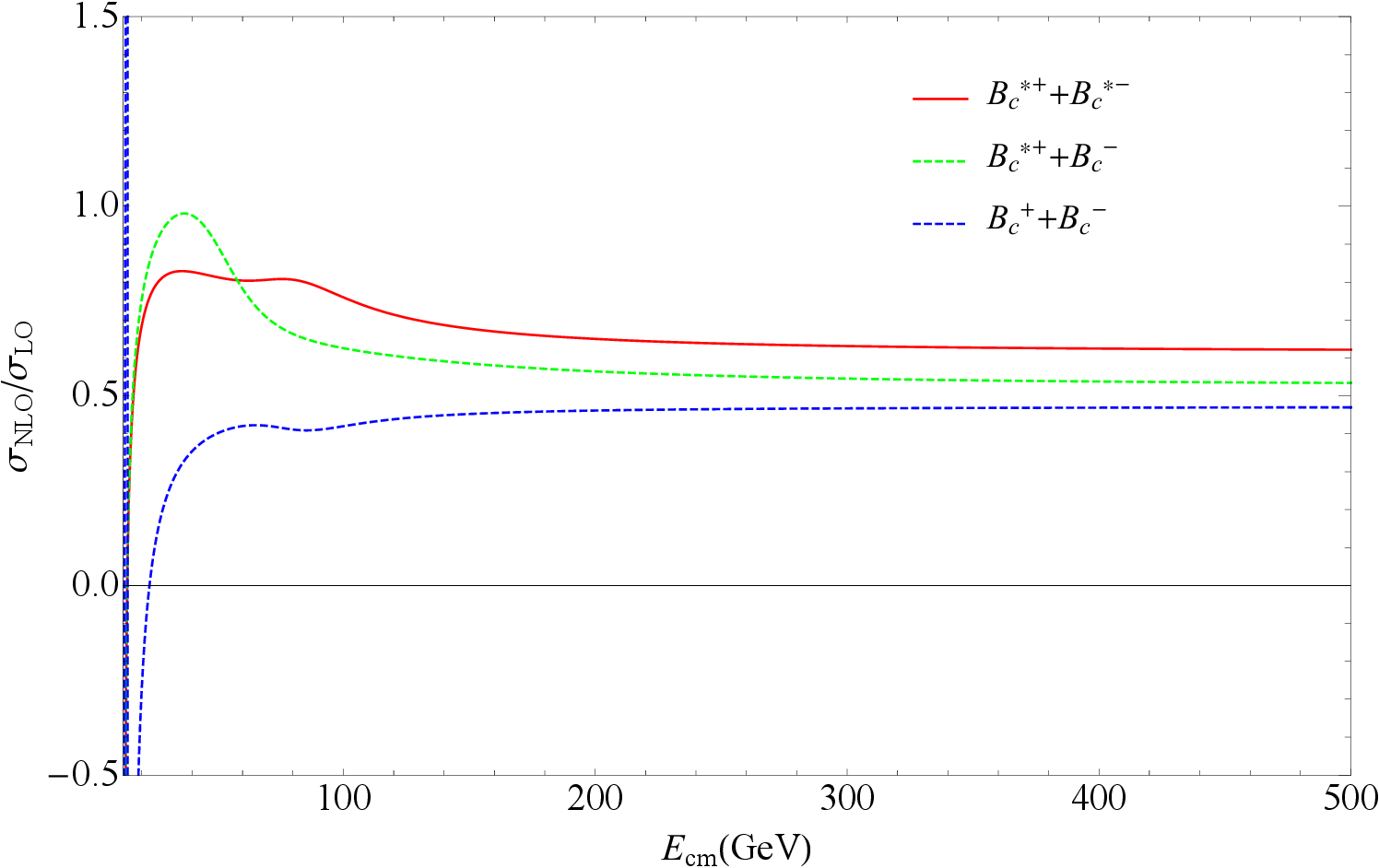}
				\includegraphics[width=0.333\textwidth]{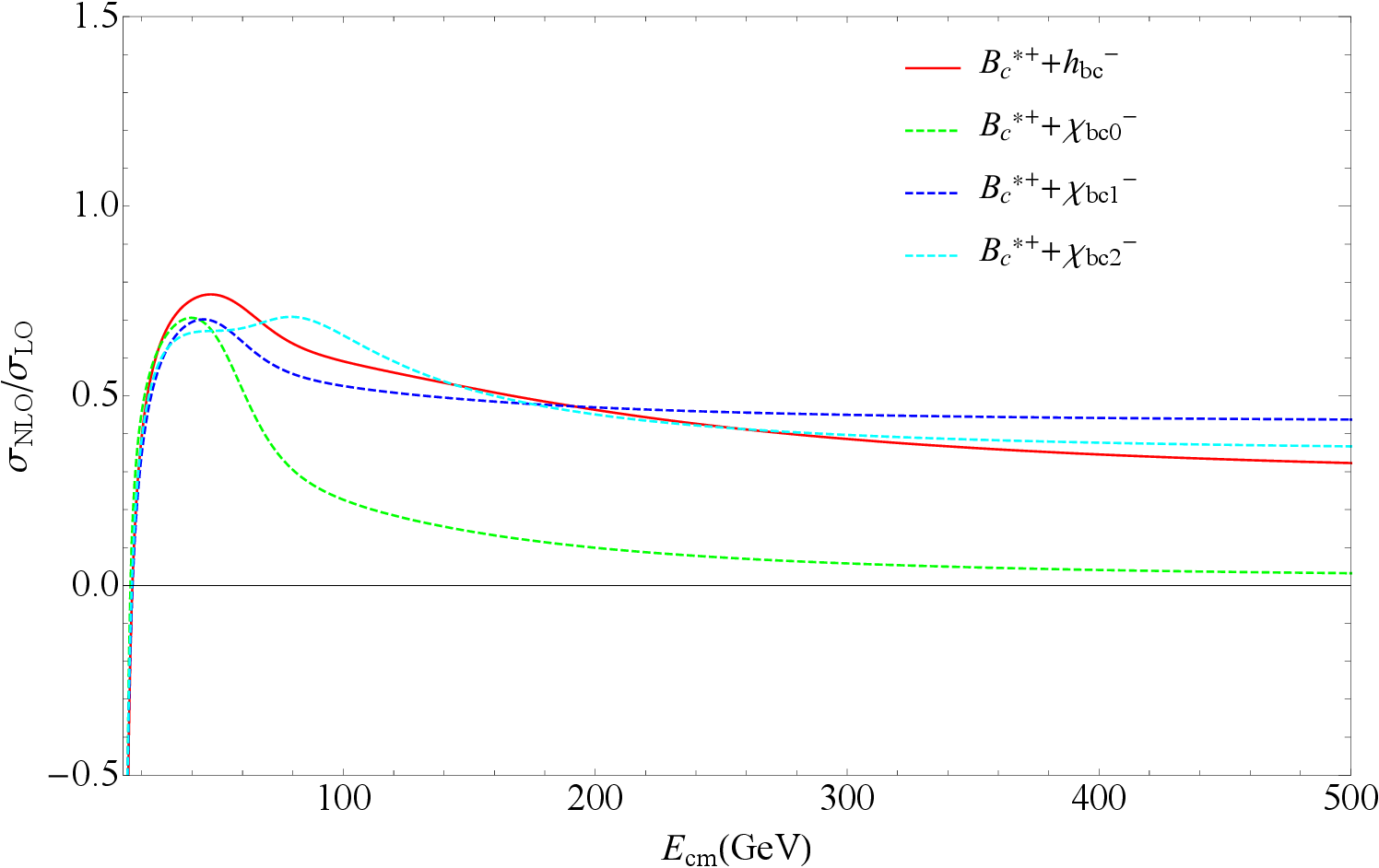}
				\includegraphics[width=0.333\textwidth]{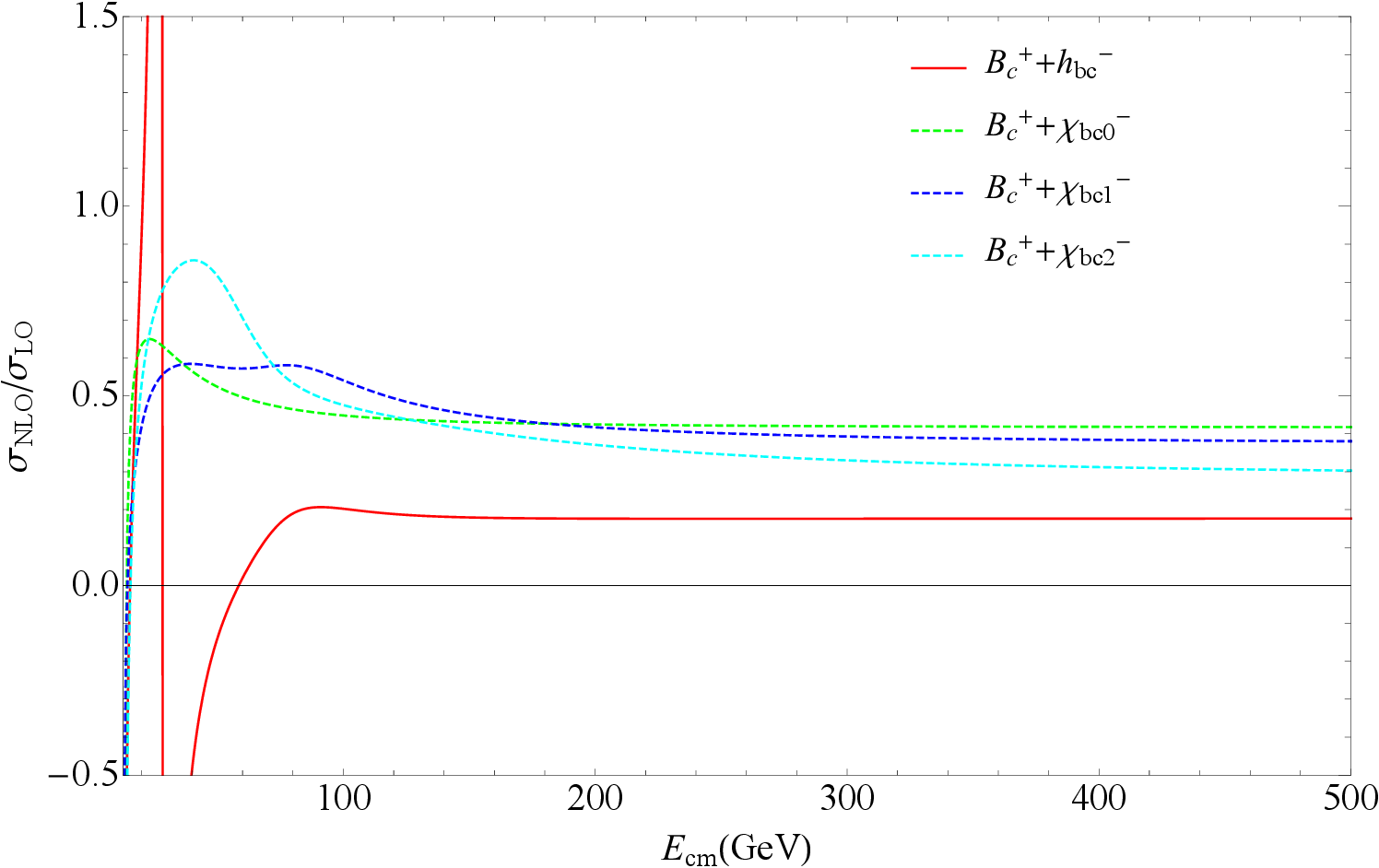}
			\end{tabular}
			\caption{(Color online) $\sigma_{NLO(v^2)}/\sigma_{LO}$ versus c.m. energy. 
The left, middle, and right panels correspond to the production of double S-wave $B_c$ meson, $B_c^{*+}+X$, and $B_c^++X$, respectively, where $X$ denotes $h_{bc}^-$ or $\chi_{bcJ}^-(J=0,1,2)$.
}
			\label{kfactor}
		\end{figure*}
	\end{widetext}

	To examine the effects of the relativistic corrections, we plot the ratio $\sigma_{\text{NLO}(v^2)}/\sigma_{\text{LO}}$ (i.e., the K-factors) as a function of the collision energy in Fig.~\ref{kfactor}. Compared to the NLO QCD cross sections, for which the K-factors are about $1.2-2.0$~\cite{Berezhnoy:2016etd}, the K-factors for the relativistic corrections are about $0.2-1.0$. At high energies, the relativistic K-factors are approximately $0.6$, indicating that the relativistic corrections reduce the $\mathcal{O}(v^0)$ cross sections by about 40\%. This level of suppression is expected and reasonable, as it lies between the $\sim50\%$ reduction calculated for double charmonia production and the $\sim20\%$ reduction for double bottomonia production\cite{Wang:2025sbx}. In the Appendix \ref{apendx}, we present the ratios of the relativistic corrections to the LO SDCs in the high-energy limit, which also reflect the contributions of the relativistic corrections.

\begin{widetext}
		\begin{figure*}[htbp]
			\begin{tabular}{c c c}
				\includegraphics[width=0.333\textwidth]{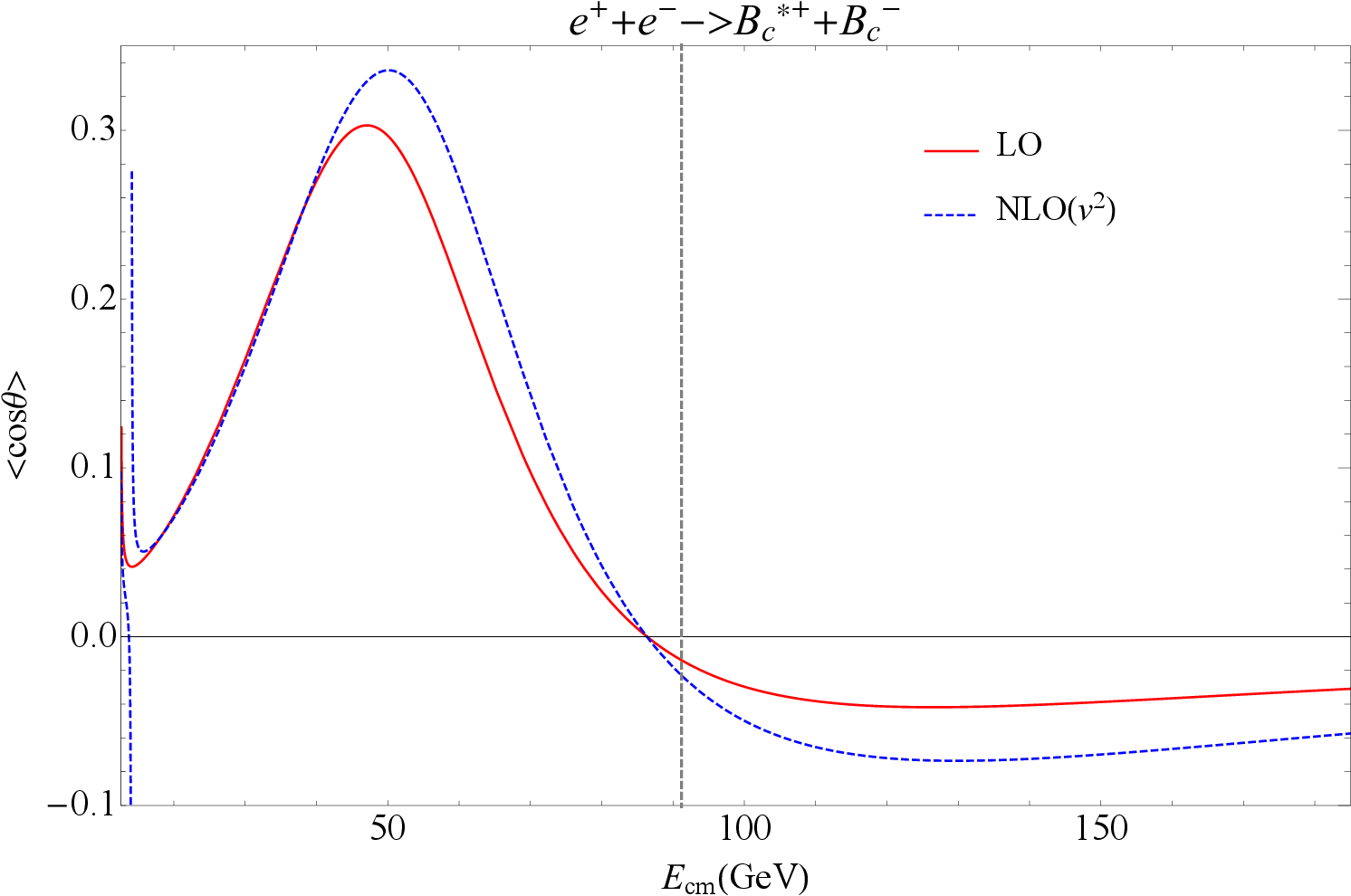}
				\includegraphics[width=0.333\textwidth]{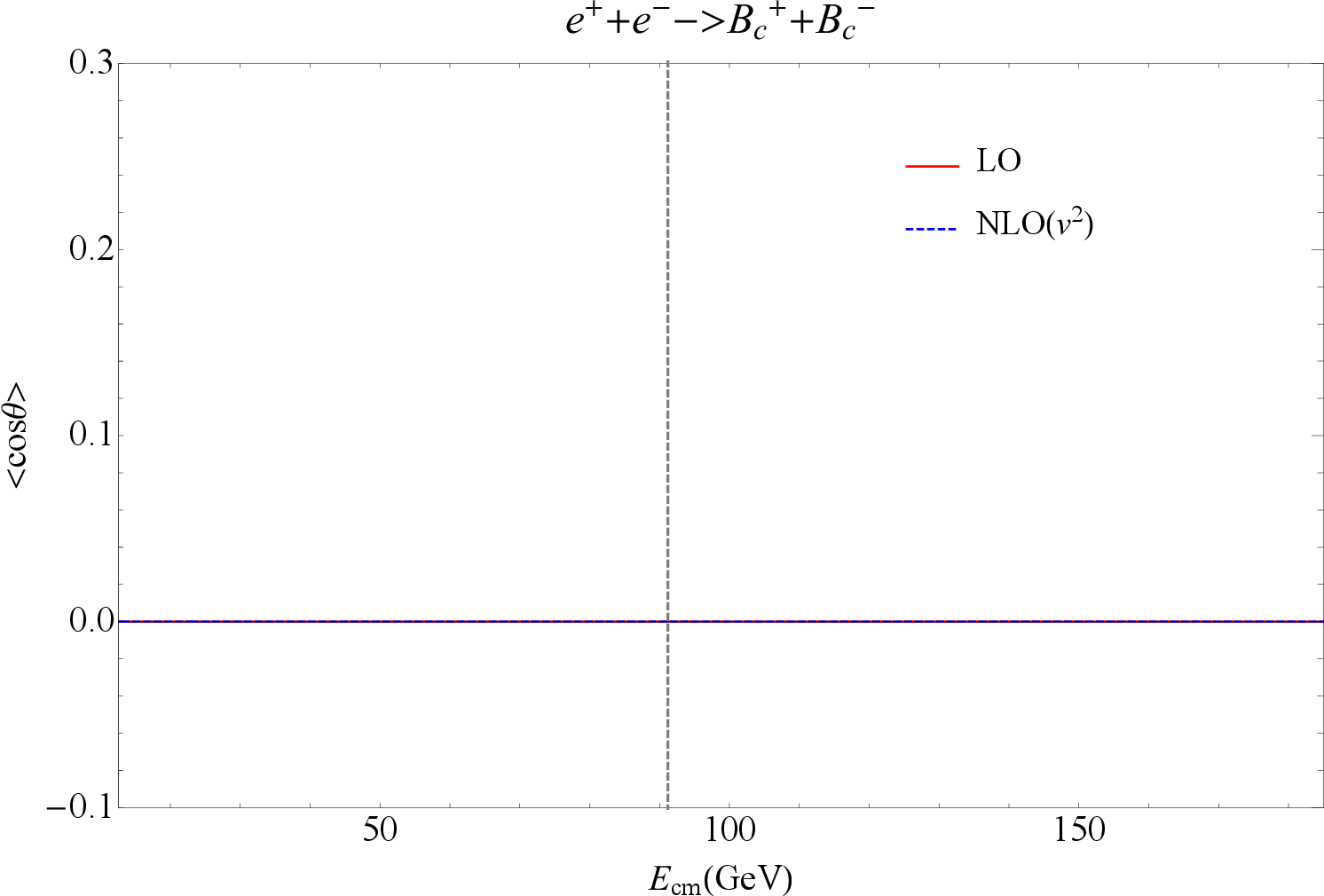}
				\includegraphics[width=0.333\textwidth]{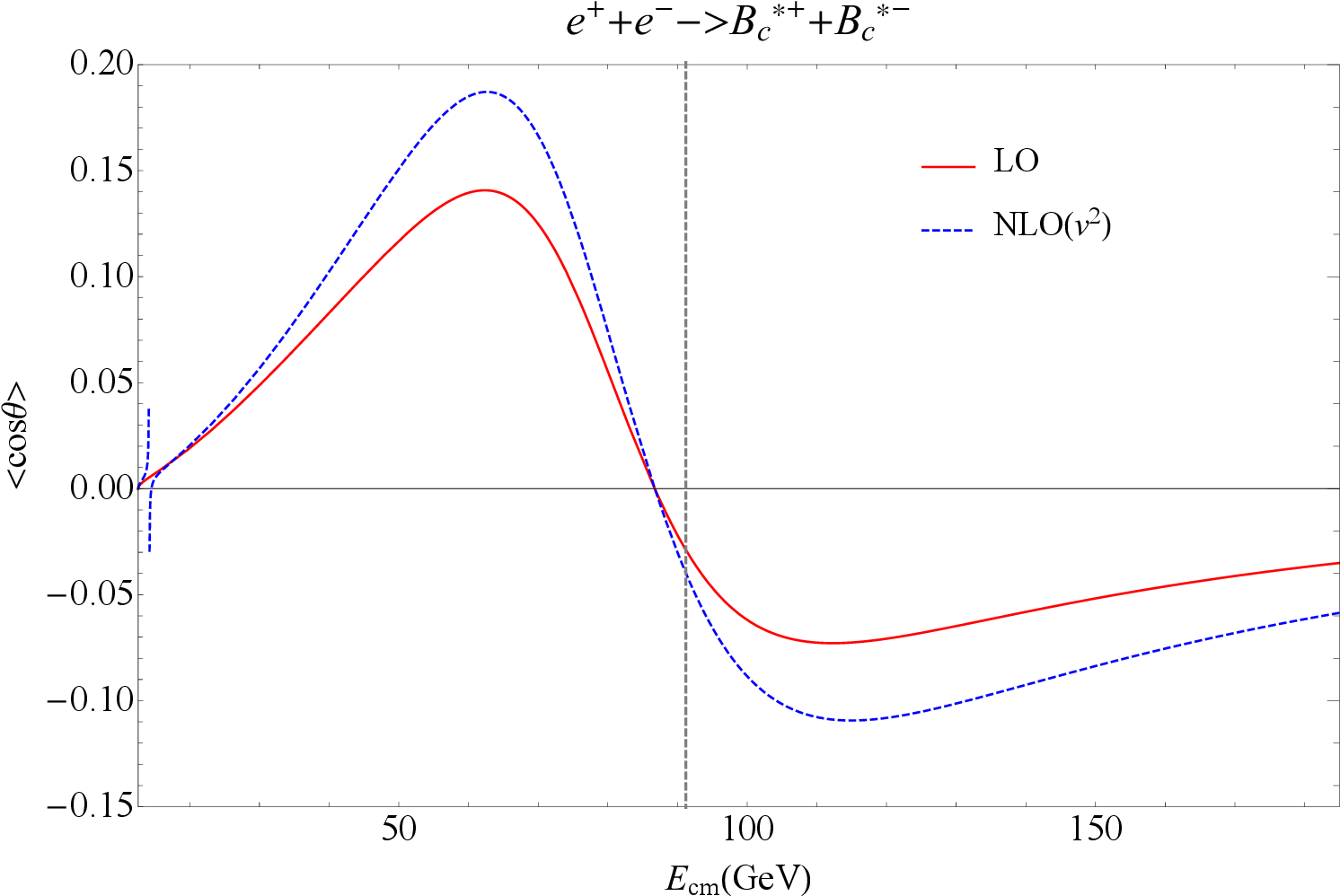}
			\end{tabular}
			
			\caption{(Color online) Azimuthal asymmetry versus c.m. energy. The red solid line represents LO and blue dashed line represents NLO($v^2$) result. The vertical dashed line marks $\sqrt{s}=m_Z$. The asymmetry is exactly zero for $B_c^++B_c^-$ production. }
			\label{asym}
		\end{figure*}
	\end{widetext}

	The production of double charmonia or double bottomonia is governed by numerous conservation laws~\cite{Wang:2025sbx,Belov:2023hpc}. For instance, C-parity conservation forbids processes like $e^+e^- \to 2J/\psi, 2\eta_c, J/\psi+h_c, \eta_c+\chi_c$ from proceeding via a $\gamma^*$ propagator. Furthermore, a given process proceeds exclusively via either the vector or the axial-vector part of the $Z^0$ current, for instance, $J/\psi+\eta_c$ production proceeds via the vector part, while $\eta_c+\chi_c$ proceeds via the axial-vector part. For charged $B_c$ mesons, these selection rules no longer hold. Consequently, we can only consider their P-parity properties.
	Since weak interactions are known to violate P-parity, we define the P-asymmetry parameter $\mathcal{A}$, following the definition in Ref.~\cite{Berezhnoy:2016etd}, as:
	\begin{align}
		\mathcal{A} = \langle \cos\theta \rangle = \frac{1}{\sigma} \int d\cos\theta \left[ \cos\theta \frac{d\sigma}{d\cos\theta} \right].
	\end{align}
	For the processes $B_c^+ + B_c^-$ and $B_c^+ + \chi_{bc0}^-$, this asymmetry is exactly zero at all energies. This is because the axial-vector part of the $Z^0$ current ($J^P=1^+$) cannot produce two $0^-$ states, and the vector part ($J^P=1^-$) cannot produce a $0^- + 0^+$ state. Therefore, there is no V-A interference and no parity violation in these specific processes. For the same reason, the process $B_c^+ + \chi_{bc0}^-$ ($0^-+0^+$) cannot proceed via a $\gamma$-propagator ($J^P=1^-$), as is also evident from Fig.~\ref{cs}.
	The asymmetry as a function of the collision energy is shown in Fig.~\ref{asym}. We find that the maximal asymmetry occurs at $\sqrt{s} \simeq m_Z/2$ and becomes nearly zero at the $Z^0$ resonance peak. Specifically, the asymmetry vanishes at the following energy points for different processes,
	\begin{align}
		\sqrt{s}(B_c^{*+}+B_c^{*-}) &= 86.74~\text{GeV}, \\
		\sqrt{s}(B_c^{*+}+B_c^{-}) &= 86.36~\text{GeV}.
	\end{align}

\begin{widetext}
		\begin{figure*}[htbp]
			\begin{tabular}{c c c}
				\includegraphics[width=0.333\textwidth]{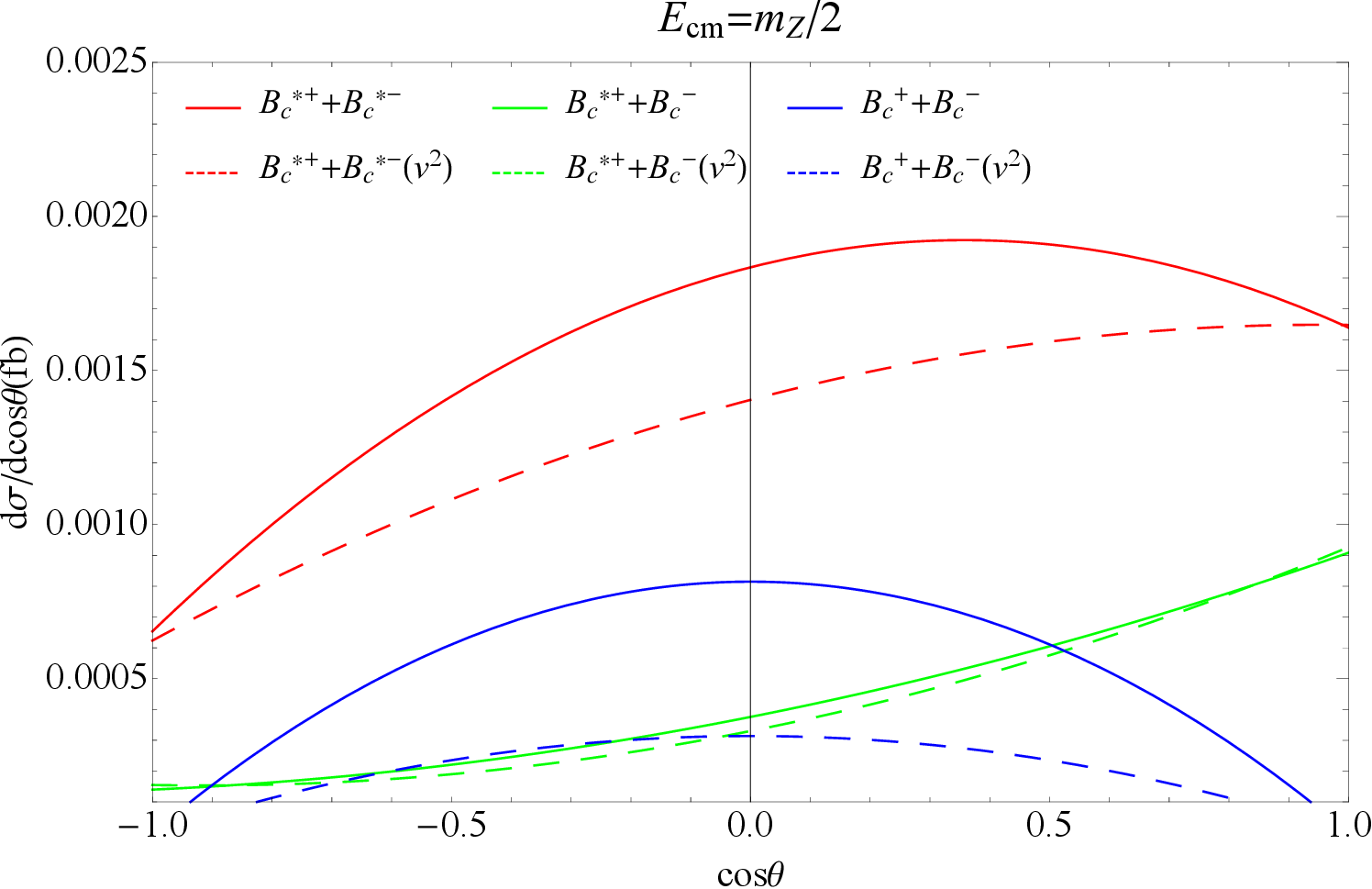}	
				\includegraphics[width=0.333\textwidth]{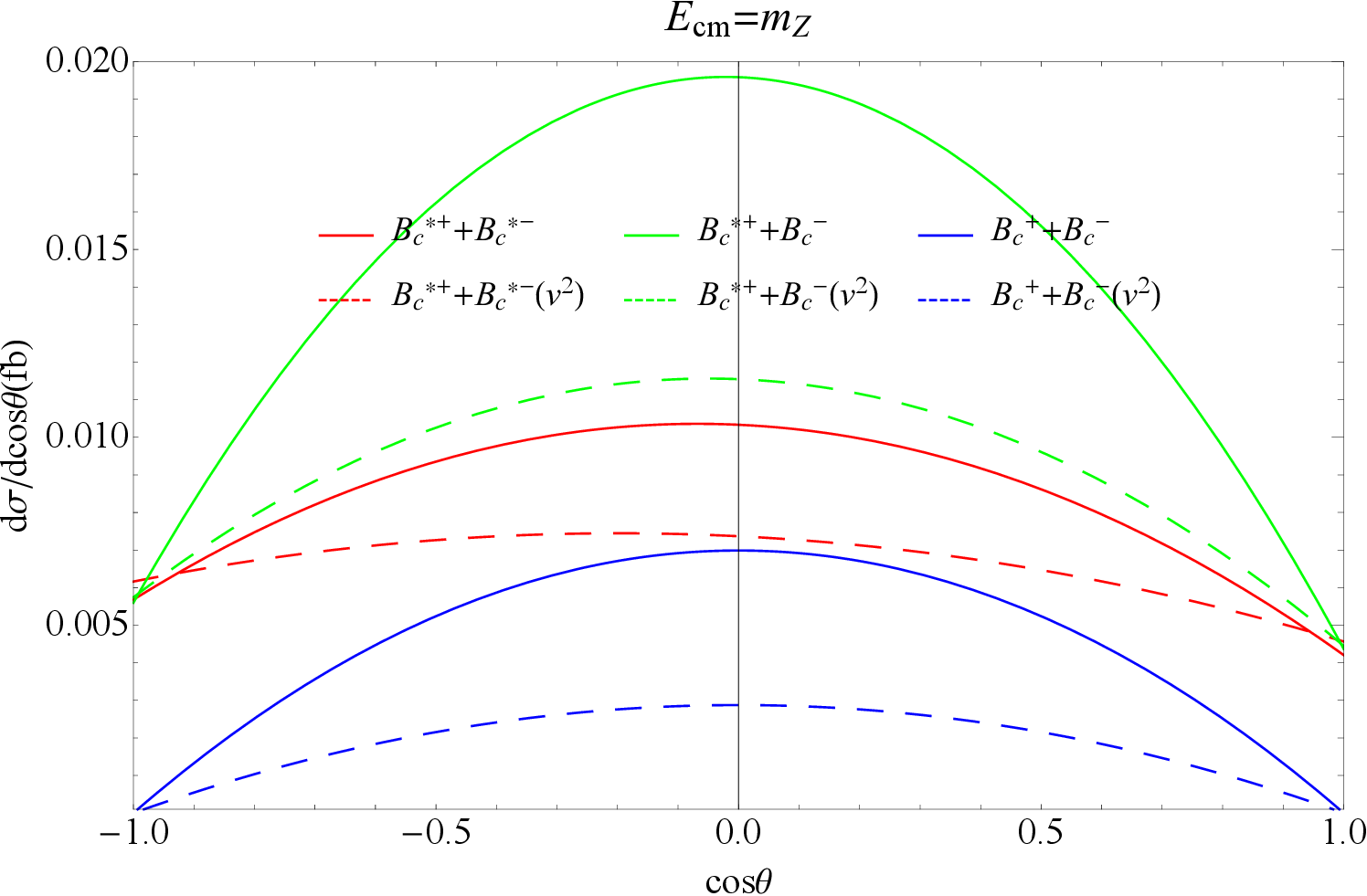}
				\includegraphics[width=0.333\textwidth]{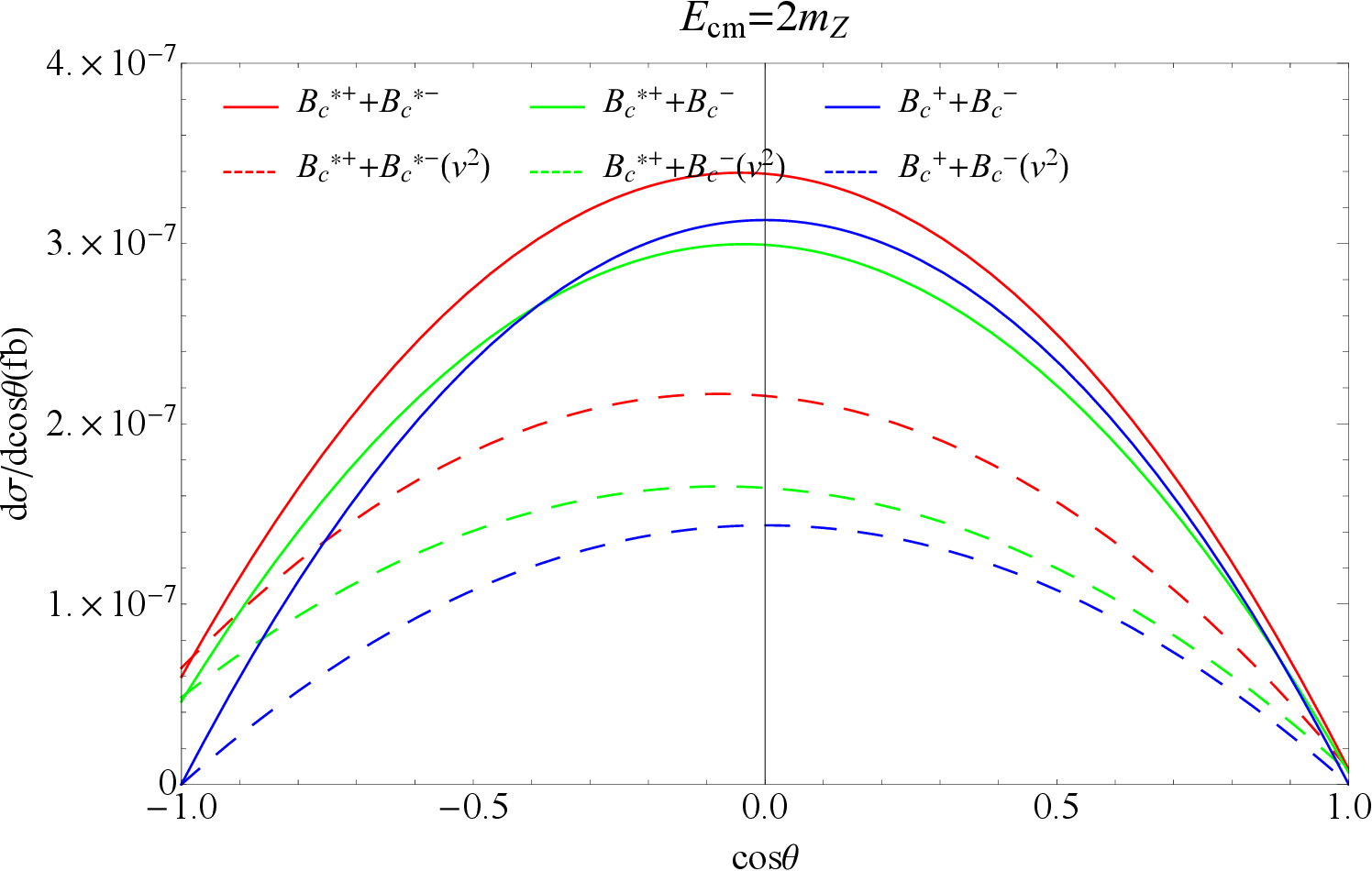}
				
			\end{tabular}
			\caption{(Color online) Differential cross section $d\sigma/d\cos\theta$ at different c.m. energy. The left, middle and right panels correspond to $\sqrt{s}=m_Z/2$, $\sqrt{s}=m_Z$, and    $\sqrt{s}=2m_Z$, respectively. The solid line represents LO results and the dashed line represents NLO($v^2$) results. }
			\label{dcos}
		\end{figure*}
	\end{widetext}

	Another notable feature is that the relativistic corrections slightly enhance the asymmetry. Moreover, we note that the azimuthal asymmetry $\langle \cos\theta \rangle$ becomes nearly zero for all production processes at sufficiently high collision energies. For instance, in $B_c^{*+} + \chi_{bcJ}^-$ production, this behavior occurs at $\sqrt{s} \geq 1000\ \text{GeV}$.
	 The angular distributions $d\sigma/d\cos\theta$ for three energy points ($\sqrt{s} = m_Z/2,\ m_Z,\ 2m_Z$) are shown in Fig.~\ref{dcos}, where $\theta$ is the angle between momenta of the electron ($k_1$) and the meson $H_1$ ($p_1$). The distributions are consistent with the azimuthal asymmetry results shown in Fig.~\ref{asym}. Specifically, the more symmetric the angular distribution, the closer the azimuthal asymmetry is to zero, and vice versa.
	
\begin{widetext}
		\begin{figure*}[htbp]
			\begin{tabular}{c c c}
				\includegraphics[width=0.333\textwidth]{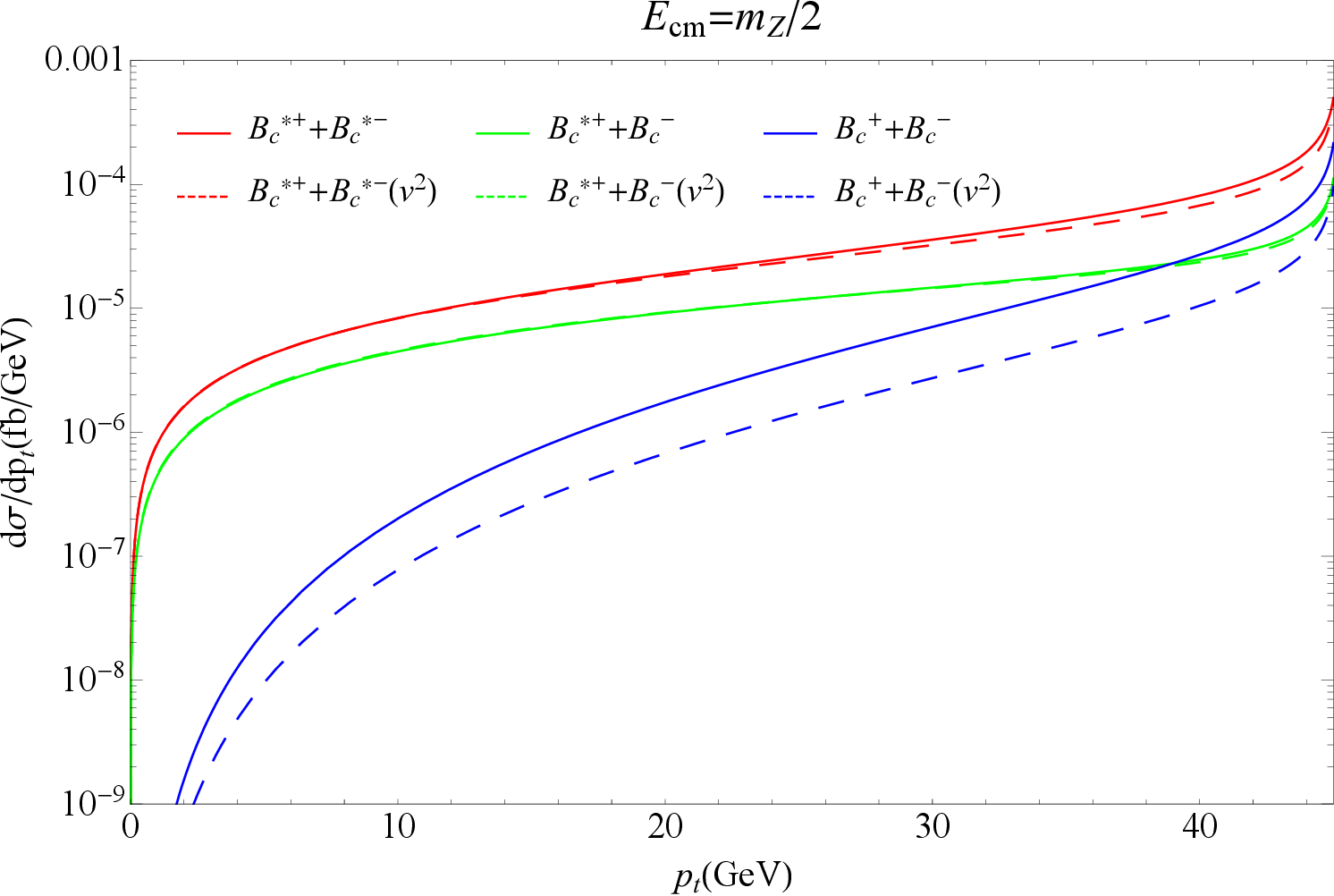}	 
				\includegraphics[width=0.333\textwidth]{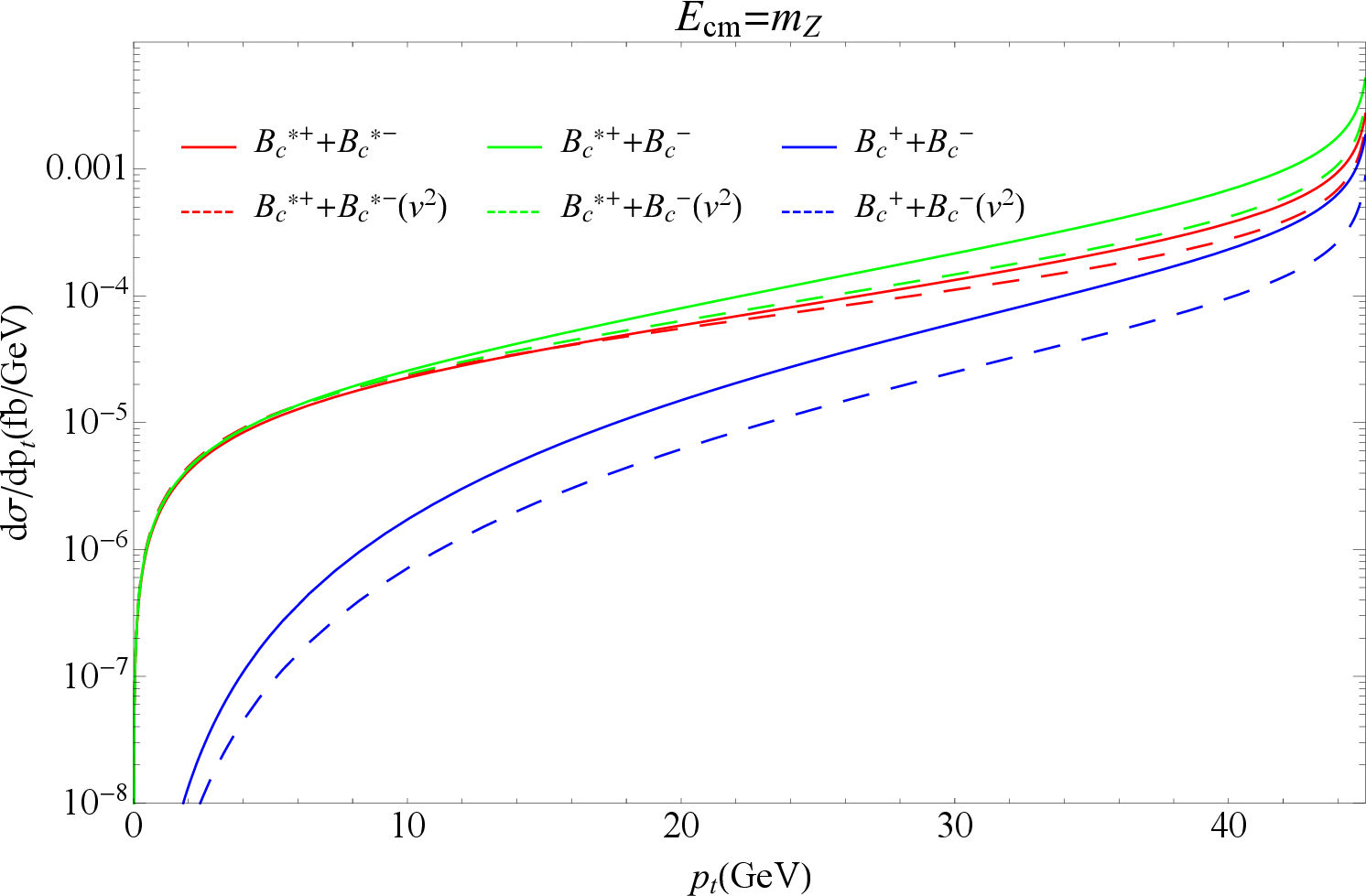}
				\includegraphics[width=0.333\textwidth]{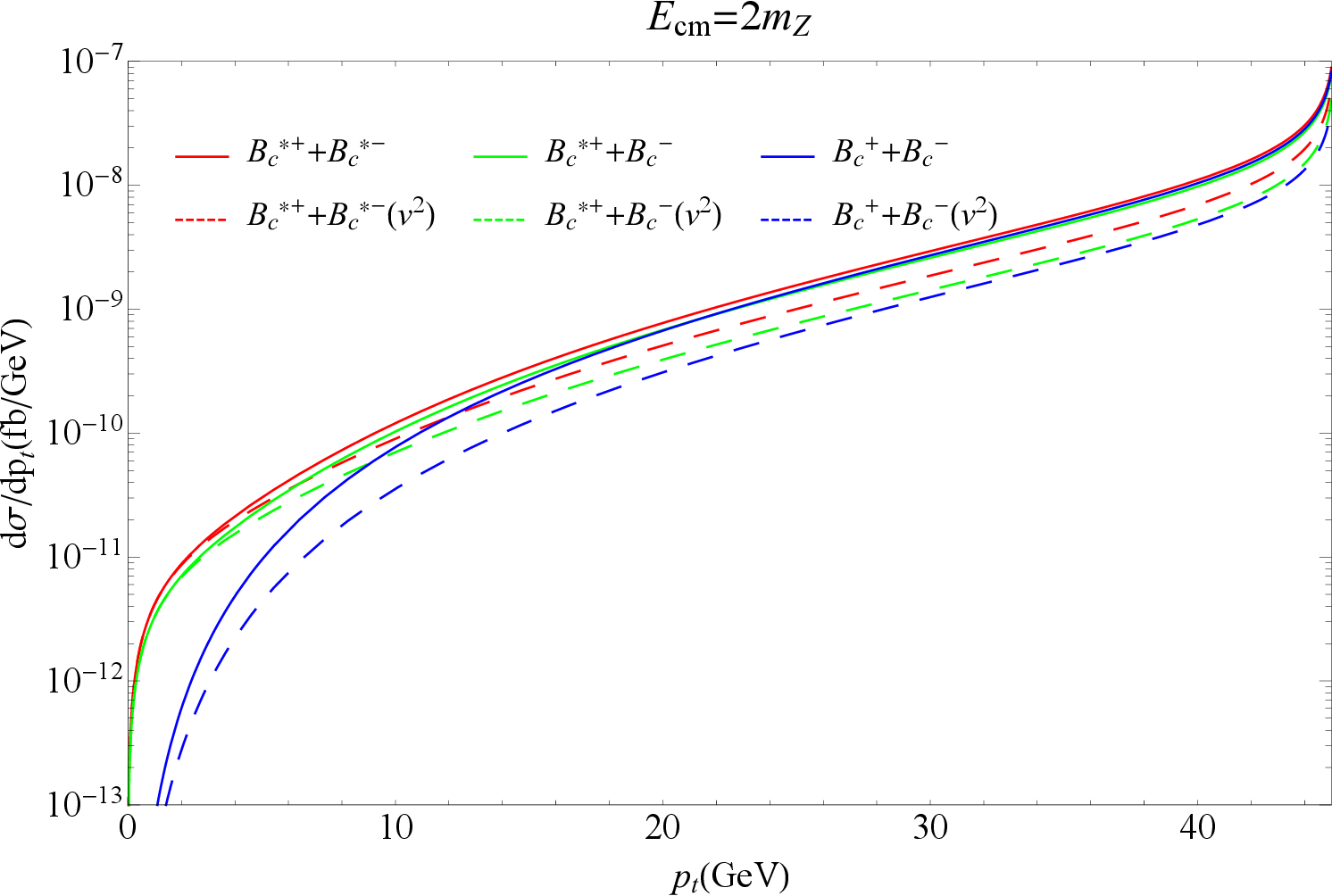}
			\end{tabular}
			\caption{(Color online) Differential cross section $d\sigma/dp_t$  at different c.m. energy. The left, middle and right panels correspond to $\sqrt{s}=m_Z/2$, $\sqrt{s}=m_Z$, and    $\sqrt{s}=2m_Z$, respectively. The solid line represents LO results and the dashed line represents NLO($v^2$) results.}
			\label{dpt}
		\end{figure*}
	\end{widetext}

The transverse momentum distribution $\frac{d\sigma}{dp_t}$ can be written as below,
	\bea
    \frac{d\sigma}{dp_t}=|\frac{d\cos\theta}{dp_t}|\frac{d\sigma}{d\cos\theta}=\frac{p_t}{|p_1|\sqrt{p_1^2-p_t^2}}\frac{d\sigma}{d\cos\theta}
	\label{pt1}
	\eea
	where the momentum of the $H_1$ meson is given by
	\bea
    p_1=\frac{\sqrt{\lambda[s,(E_{c}+E_{\bar{b}})^2,(E_{b}+E_{\bar{c}})^2]}}{2\sqrt{s}},~~\lambda[x,y,z]= x^2+y^2+z^2-2(xy+yz+zx) 
	\label{pt2}
	\eea
	Combining Eqs. (\ref{pt1}) and (\ref{pt2}), we can also get an $\mathcal{O}(v^2)$  expression,
\bea
	\frac{d\sigma}{dp_t}= \frac{4 p_t [s (1 - 4 r) (1 - 4 r - 4 r_{p_t}) + 16 m_b m_c (1 - 4 r - 2 r_{p_t}) v^2]}{s^2 (1 - 4 r)^{3/2} (1 - 4 r - 4 r_{p_t})^{3/2}} \frac{d\sigma}{d\cos\theta}+\mathcal{O}(v^4), ~~r_{p_t}=\frac{p_t^2}{s}
	\eea
	Using the above expressions, we show the transverse momentum distributions $\mathrm{d}\sigma/\mathrm{d}p_t$ at different c.m. energy points in Fig.~\ref{dpt}.

	\begin{table}
		\caption{Production  cross sections at $\sqrt{s}=m_Z/2, m_Z, 2m_Z$ for different renormalization scales. LO and NLO($v^2$) means leading order and next-to-leading order in the $v^2$ expansions, respectively.  The K-factor, defined as $K=\sigma_{NLO(v^2)}/\sigma_{LO}$, is identical across different scales $\mu$.  The results of $B_c^++\chi_{bc0}^-$ at $\sqrt{s}=m_Z/2$ are not shown because their cross sections ($\sigma\sim10^{-7}fb$) are much smaller than those of other processes. }
		\begin{tabular}{|c|c|c||c|c||c|c||c|c||c|}
			\hline
			~	&\multicolumn{2}{c||}{$\mu=\sqrt{s}/2$} &  \multicolumn{2}{c||}{$\mu=\sqrt{s}$} 	&\multicolumn{2}{c||}{$\mu=2\sqrt{s}$} &   \multicolumn{2}{c||}{$\mu=2M_{B_c}$}&K\\
			\hline
			~&~LO~&NLO($v^2$) & ~LO~&NLO($v^2$)& LO~&NLO($v^2$) & ~LO~&NLO($v^2$)&\\
			\hline
			\hline
		$	\sqrt{s}=m_Z/2$&\multicolumn{9}{c|}{units:~$\times 10^{-2}fb$} \\
		\hline
		$B_c^{*+}+B_c^{*-}$& 0.428 & 0.351   &0.321   &0.263 & 0.249   &0.204  &  0.569& 0.466 &0.82 \\
		\hline
		$B_c^{*+}+B_c^{-}$& 0.113& 0.107   &0.085   &0.080 & 0.066   & 0.062 & 0.151 & 0.142 &0.94 \\
		\hline
		$B_c^{+}+B_c^{-}$&0.145&  0.056  & 0.109  & 0.042& 0.084   & 0.032 &0.193  & 0.074 & 0.39\\
		\hline
		$B_c^{*+}+h_{bc}^{-}$&0.044& 0.034   & 0.033  &0.026 &  0.026  & 0.020 & 0.059 & 0.045 &0.77 \\
		\hline
		$B_c^{*+}+\chi_{bc0}^{-}$&0.033& 0.022   & 0.025  &0.016 &   0.019 &0.013  & 0.044 &  0.029&0.67 \\
		\hline
		$B_c^{*+}+\chi_{bc1}^{-}$&0.032& 0.022   & 0.024  &0.017 & 0.019   &0.013  &0.042  &  0.030& 0.70\\
		\hline
		$B_c^{*+}+\chi_{bc2}^{-}$&0.060& 0.040   &0.045   &0.030 &  0.035  & 0.023 & 0.080 &0.053  &0.67 \\
		\hline
		$B_c^{+}+h_{bc}^{-}$&0.0028& -0.0007   & 0.0021  & -0.0005& 0.0016   & -0.0004 & 0.0037 &  -0.0009&-0.25 \\
		\hline
		$B_c^{+}+\chi_{bc0}^{-}$&  &    &   & &    &  &  &  &0.54 \\
		\hline
		$B_c^{+}+\chi_{bc1}^{-}$& 0.083& 0.048   &0.062   & 0.036& 0.048   &0.028  &0.011  &0.064  &0.58 \\
		\hline
		$B_c^{+}+\chi_{bc2}^{-}$&0.024& 0.020   &0.018   &0.015 & 0.014   & 0.012 & 0.032 & 0.027 & 0.85\\
		\hline	
			\hline
			$	\sqrt{s}=m_Z$&\multicolumn{9}{c|}{units:~$\times10^{-1}fb$} \\
			\hline
			$B_c^{*+}+B_c^{*-}$&0.220 &0.173   & 0.171   &0.134 & 0.136   &0.107  &0.390  & 0.306& 0.78\\
			\hline
			$B_c^{*+}+B_c^{-}$&0.379& 0.242  &0.294    &0.188 & 0.235   & 0.150 & 0.672 & 0.429 &0.64\\
			\hline
			$B_c^{+}+B_c^{-}$& 0.120&  0.049 & 0.093   & 0.038&  0.074  &0.031  & 0.213 & 0.087& 0.41\\
			\hline
			$B_c^{*+}+h_{bc}^{-}$& 0.048& 0.029  &0.037    &0.022 & 0.030   &  0.018& 0.084 &0.051 &0.61 \\
			\hline
			$B_c^{*+}+\chi_{bc0}^{-}$&0.025&  0.018 & 0.019   &0.014 & 0.015   &0.011  & 0.044 & 0.032 &0.74\\
			\hline
			$B_c^{*+}+\chi_{bc1}^{-}$&0.072 & 0.039  & 0.056   &0.030 & 0.045   & 0.024 & 0.128 & 0.069 &0.54\\
			\hline
			$B_c^{*+}+\chi_{bc2}^{-}$& 0.038& 0.026  & 0.029   &0.020 & 0.024   &0.016  &0.067  &0.046 &0.69 \\
			\hline
			$B_c^{+}+h_{bc}^{-}$&0.008 &0.002   &0.006    &0.001 &0.005    & 0.001 & 0.014 & 0.003 &0.21\\
			\hline
			$B_c^{+}+\chi_{bc0}^{-}$&0.0005 & 0.0002  &0.0004    &0.0002 &  0.0003  &  0.0001&  0.0009& 0.0004 &0.45\\
			\hline
			$B_c^{+}+\chi_{bc1}^{-}$& 0.049&  0.027 & 0.038   &0.021 &0.030    &0.017  & 0.086 & 0.048 &0.56\\
			\hline
			$B_c^{+}+\chi_{bc2}^{-}$&0.048 & 0.024  & 0.037   &0.018 & 0.029   &0.015  &0.084  &0.042 & 0.50\\
			\hline	
			\hline
			$	\sqrt{s}=2m_Z$&\multicolumn{9}{c|}{units:~$\times10^{-6}fb$} \\
			\hline
			$B_c^{*+}+B_c^{*-}$& 0.594& 0.390  &0.474    &0.312 & 0.387   &0.255  & 1.355 & 0.890 &0.66\\
			\hline
			$B_c^{*+}+B_c^{-}$& 0.522&  0.298 & 0.417   &0.238 & 0.340   & 0.194 & 1.190 &  0.679&0.57\\
			\hline
			$B_c^{+}+B_c^{-}$&0.673 & 0.309  &0.523    &0.240 & 0.417   &0.192  & 1.192 &0.547 &0.46 \\
			\hline
			$B_c^{*+}+h_{bc}^{-}$& 0.046&  0.022 &0.037    &0.018 &  0.030  &0.014  & 0.105 &0.050 &0.48 \\
			\hline
			$B_c^{*+}+\chi_{bc0}^{-}$&0.015& 0.011  & 0.012   &0.008 &  0.010  &  0.007& 0.034 & 0.025& 0.73\\
			\hline
			$B_c^{*+}+\chi_{bc1}^{-}$& 0.096& 0.046  & 0.077   &0.036 & 0.063   &0.030  & 0.219 & 0.104&0.48 \\
			\hline
			$B_c^{*+}+\chi_{bc2}^{-}$& 0.067& 0.032  & 0.054   & 0.025& 0.044   &0.021  & 0.153 & 0.072&0.47 \\
			\hline
			$B_c^{+}+h_{bc}^{-}$&0.027& 0.005  &0.022    &0.004 & 0.018   & 0.003 & 0.062 & 0.011 &0.18\\
			\hline
			$B_c^{+}+\chi_{bc0}^{-}$&0.0007 & 0.0003  & 0.0006   & 0.0003& 0.0005   & 0.0002 &0.0017  &0.0007  &0.43\\
			\hline
			$B_c^{+}+\chi_{bc1}^{-}$&0.110 &  0.047 &  0.088  &0.038 & 0.072   & 0.031 &0.252  & 0.107& 0.43\\
			\hline
			$B_c^{+}+\chi_{bc2}^{-}$&0.060 & 0.023  &0.048    &0.018 & 0.039   &0.015  & 0.137 & 0.053 &0.38\\
			\hline	
		\end{tabular}
		\label{TCS1}
	\end{table}
	\FloatBarrier

	\subsection{CO contributions}
	
CO contributions are expected to be significant in annihilation and hadroproduction processes\cite{Wu:2002ig,Chang:2005bf}, but relatively small in electron-positron collisions \cite{Yang:2011ps,Wang:2025sbx} and some indirect production mechanisms\cite{Chang:2007si, Yang:2010yg}. 
Contributions of CO states are significant in high energy region for double heavy quarkonium production in electron-positron collisions, where the gluon fragmentation process is most important \cite{Wang:2025sbx}. In contrast, there are no gluon fragmentation diagrams for the production of paired $B_c$ mesons.
	
The CO channels actually represent the inclusive processes for the soft gluon emission or absorption. The complete calculations of LO inclusive production invoking $CS+CO+g,CO+CO+g$ which are of the NLO of $\alpha_s$ compared with the CS channels. 
For CO pair channels in a special case, in which the two final states with an energy close to half of the c.m. energy recoil without the emission of additional light particles, those contribute to the exclusive CS double mesons production in the experimental measurements. This specific configuration leads to the breakdown of factorization due to nonperturbative effects which, fortunately, are suppressed by the inverse powers of the collision energy and are usually negligible. 
To obtain a rough estimate of these contributions, we follow the velocity scaling rules of NRQCD and the approach used in Refs. \cite{Wu:2002ig,Chang:2005bf,Chang:2007si,Wu:2008cn,Yang:2010yg,Yang:2011ps}, assuming a relation between the CO and CS LDMEs,
	\begin{equation}
		\langle\mathcal{O}^H_8\rangle \simeq \Delta_S(v)^2 \langle\mathcal{O}^H_1\rangle
	\end{equation}
	Here, $\Delta_S(v)$ is of order $v^2$, with typical values $v^2 \sim 0.1$–0.3, and $\Delta_S(v) \sim \alpha_s(M_{B_c} v)$.
	
	We now estimate the CO contribution to the cross section at the $Z^0$ peak. As noted previously, the electroweak contributions are several orders of magnitude smaller than the QCD contributions, so it suffices to consider the $\gamma/Z$–$g$ type processes. 
The color factor for the CS amplitude squared is $|C_s|^2 = C_F^2$, while for the CO one it is $|C_o|^2 = C_F/(2N_c)$.
The CO cross section is related to the CS cross section by the below expression,
	\begin{equation}
		\sigma_8 = \sigma_1 \times \frac{|C_o|^2}{|C_s|^2} \times v^8 = \sigma_1 \times \frac{v^8}{8}
	\end{equation}
	
	Taking $v^2 = 0.15$ and considering $B_c^{*+} + B_c^{-}$ production (via ${}^3S_1^{[1]} + {}^1S_0^{[1]}$) as an example, the CS cross section at the $Z^0$ peak is 0.030 fb. The four main CO channels are:
	${}^3S_1^{[8]} + {}^1S_0^{[8]}$, ${}^3S_1^{[8]} + {}^3S_1^{[8]}$, ${}^1S_0^{[8]} + {}^3S_1^{[8]}$, and ${}^1S_0^{[8]} + {}^1S_0^{[8]}$. 
	The total CO cross section is then estimated as:
	\begin{equation}
		\sigma_8 \simeq 4 \times \sigma_1 \times \frac{v^8}{8} \simeq 7.6 \times 10^{-6} \ \mathrm{fb}
	\end{equation}
	This is approximately three orders of magnitude smaller than the CS contribution.

	\subsection{Uncertainties}
	Theoretical uncertainties in our calculation arise from several sources, including the LDMEs, heavy quark masses, and the choice of renormalization scale.
	
	For the LDMEs, various studies\cite{Ebert:2011jc,Liao:2014rca,Eichten:2019gig,Berezhnoy:2019jjs,Berezhnoy:2021wrc,Chen:2024dkx} report values in the ranges $|R_S(0)|^2 \in (1.30, 6.21)$ and $|R'_P(0)|^2 \in (0.16, 0.90)$. Based on these ranges, we estimate that the LDMEs can introduce uncertainties of approximately -57\% to +870\% for double S-wave $B_c$ production, and -66\% to +809\% for ``S+P" waves $B_c$ pair production.
	
	The renormalization scale uncertainty is partially illustrated in Table~\ref{TCS1}. At $\sqrt{s} = m_Z$, the dependence on the scale is similar to that observed in double heavy quarkonium production (cf. Figs.~10 and 11 in Ref.~\cite{Wang:2025sbx}): it decreases with increasing collision energy and can lead to variations of about -20\% (for $\mu=2m_Z$) to +128\% (for $\mu=2M_{B_c}$). For simplicity, these are not explicitly shown in the figures.
	
	To quantify the uncertainties from the heavy quark masses, we adopt $m_c = (1.5 \pm 0.3)$ GeV and $m_b = (4.8 \pm 0.4)$ GeV, following Refs.~\cite{Deng:2010aq,Yang:2010yg,Yang:2011ps}. The resulting total cross sections at the $Z^0$ pole are presented in Tables~\ref{QMU1} and \ref{QMU2}. The cross sections are more sensitive to the charm quark mass, which is consistent with previous findings\cite{Deng:2010aq,Yang:2010yg,Yang:2011ps}. We also find that the cross sections increase as the charm quark mass decreases, but decrease as the bottom quark mass decreases—a behavior similar to that seen for the $(c\bar{b})_1[{}^3S_1,{}^3P_2]$ states in Refs.~\cite{Yang:2010yg,Yang:2011ps}. In general, the uncertainty caused by the $\pm0.3 GeV$ variation of   charm quark mass is $-40\%\sim+100\%$ for double S-wave $B_c$ meson production and $-60\%\sim+200\%$ for P-wave $B_c$ meson-involved production processes,   while that caused by the $\pm0.4 GeV$ variation of   bottom quark mass is $-10\%\sim+10\%$ for double S-wave $B_c$ meson production and $-20\%\sim+20\%$ for P-wave $B_c$ meson-involved production processes.

	\begin{table}
		\caption{Uncertainties for the total cross sections at $\sqrt{s}=m_Z$ with varying $m_c$, where $m_b$ is fixed to be 4.8 GeV.  }
		\begin{tabular}{|c|c|c|c|c|c|c| }
			\hline
			$\sigma~(\times10^{-1}fb)$	&\multicolumn{2}{c|}{$m_c=1.2~GeV$} &  \multicolumn{2}{c|}{$m_c=1.5~GeV$} 	&\multicolumn{2}{c|}{$m_c=1.8~GeV$} \\
			\hline
			&~LO~&NLO($v^2$) & ~LO~&NLO($v^2$)& LO~&NLO($v^2$) \\
			\hline
			$B_c^{*+}+B_c^{*-}$&0.425  &0.311 & 0.171  &0.134  & 	  0.085 & 0.071 \\
			\hline
			$B_c^{*+}+B_c^{-}$&  0.649& 0.373  &0.294   &0.188  &	0.162    & 0.113  \\
			\hline
			$B_c^{+}+B_c^{-}$& 0.198  &0.066 &  0.093 & 0.038 &	 0.052  & 0.025  \\
			\hline
			$B_c^{*+}+h_{bc}^{-}$&0.149  &0.080 &0.037   & 0.022 &	 0.012  &0.008  \\
			\hline
			$B_c^{*+}+\chi_{bc0}^{-}$&0.073  &0.046 & 0.019  &0.014  &	0.007   &0.006  \\
			\hline
			$B_c^{*+}+\chi_{bc1}^{-}$&  0.183  & 0.084 &0.056   &0.030  &  0.023  &0.014 \\
			\hline
			$B_c^{*+}+\chi_{bc2}^{-}$&0.125   &0.077  & 0.029  & 0.020 & 0.009   &0.007 \\
			\hline
			$B_c^{+}+h_{bc}^{-}$&   0.0194  & 0.0021 & 0.0061  &0.0013  &  0.0024  &0.0007 \\
			\hline
			$B_c^{+}+\chi_{bc0}^{-}$&   0.0021 & 0.0007 & 0.0004  &0.0002  & 0.00008   & 0.00005\\
			\hline
			$B_c^{+}+\chi_{bc1}^{-}$& 0.134  &0.066  &0.038   & 0.021 & 0.014   &0.009 \\
			\hline
			$B_c^{+}+\chi_{bc2}^{-}$& 0.142  &0.061  &0.037   &0.018  & 0.012   &0.007 \\
			\hline
		\end{tabular}
		\label{QMU1}
	\end{table}
	
	\begin{table}
		\caption{Uncertainties for the total cross sections at $\sqrt{s}=m_Z$  with varying $m_b$, where $m_c$ is fixed to be 1.5 GeV.  }
		\begin{tabular}{|c|c|c|c|c|c|c| }
			\hline
			$\sigma~(\times10^{-1}fb)$	&\multicolumn{2}{c|}{$m_b=4.4~GeV$} &  \multicolumn{2}{c|}{$m_b=~4.8GeV$} 	&\multicolumn{2}{c|}{$m_b=5.2~GeV$} \\
			\hline
			&~LO~&NLO($v^2$) & ~LO~&NLO($v^2$)& LO~&NLO($v^2$) \\
			\hline
			$B_c^{*+}+B_c^{*-}$&0.137   &0.108  & 0.171  &0.134  & 0.215	   &0.168   \\
			\hline
			$B_c^{*+}+B_c^{-}$& 0.254  & 0.166 & 0.294  &0.188  &  0.344  &  0.216 \\
			\hline
			$B_c^{+}+B_c^{-}$&0.085   &0.038 & 0.093  &0.038  &	0.102   & 0.039  \\
			\hline
			$B_c^{*+}+h_{bc}^{-}$& 0.028 &0.017 &  0.037 &0.022  &	 0.049  &0.029  \\
			\hline
			$B_c^{*+}+\chi_{bc0}^{-}$&0.014  &0.011 &  0.019 & 0.014 &0.026	   &0.018  \\
			\hline
			$B_c^{*+}+\chi_{bc1}^{-}$& 0.049  & 0.028 & 0.056  & 0.030 & 0.065   &0.034 \\
			\hline
			$B_c^{*+}+\chi_{bc2}^{-}$& 0.022   &0.015  & 0.029  &0.020  & 0.039   &0.027 \\
			\hline
			$B_c^{+}+h_{bc}^{-}$&   0.0057 &0.0014  &0.0061   & 0.0013 &  0.0065  & 0.0011\\
			\hline
			$B_c^{+}+\chi_{bc0}^{-}$& 0.00026   &0.00013  &0.00038   & 0.00017 & 0.00052   &0.00021 \\
			\hline
			$B_c^{+}+\chi_{bc1}^{-}$&   0.031  &0.018  &0.038   & 0.021 & 0.047   &0.026 \\
			\hline
			$B_c^{+}+\chi_{bc2}^{-}$&  0.030   &0.015  &0.037   &0.018  & 0.046   &0.022 \\
			\hline
		\end{tabular}
		\label{QMU2}
	\end{table}

	\section{Conclusions}
	\label{conclusion}
	
	Based on previous studies of $B_c$ meson pair production in $e^+e^-$ collisions, including LO calculation \cite{Liao_2023,liaoqili,liaoqili2}, NLO QCD correction\cite{Berezhnoy:2016etd}, and relativistic corrections within the RQM to both s-channel\cite{Karyasov:2016hfm} and t-channel double-photon processes\cite{Berezhnoy:2019jjs}, we have investigated the relativistic corrections to double $B_c$ production in $e^+e^-$ annihilation within the NRQCD factorization framework\cite{Bodwin:1994jh},   calculating both $\gamma^*$- and $Z^0$-propagated processes, with c.m. energies ranging from the production threshold $2M_{B_c}$ up to $2m_Z$.
	
	The results demonstrate that the relativistic corrections are substantial, with K-factors of approximately 0.6, corresponding to a ~40\% reduction of the LO cross sections. This effect is comparable in magnitude to the NLO QCD corrections, which have K-factors of about 1.4 \cite{Berezhnoy:2016etd}. We  also present the azimuthal asymmetry arising from P-parity violation as a function of the c.m. energy, along with the differential cross sections (angular distribution and transverse momentum distribution) at specific energy points ($m_Z/2$, $m_Z$, and $2m_Z$). And the theoretical uncertainties from quark masses and the renormalization scale are discussed.
	
	Considering the negligible contributions from the t-channel double-photon mechanism and CO states, the small reconstruction efficiency or decay branching fractions (e.g., $\mathcal{B}(B_c^{\pm} \to J/\psi \pi^{\pm}) \approx 0.5\%$ \cite{Chang:1992pt}), and the significant negative relativistic corrections, despite the enhancing effect of NLO QCD corrections\cite{Berezhnoy:2016etd}, the observation of double $B_c$ production shall be challenging (as also pointed out in Ref. \cite{Wei:2018xlr}) for future $ee$ collider $Z$-factory experiments.
	
	Therefore, alternative production modes may be more promising. These include operation near the production threshold ($\sqrt{s} \sim 15\ \mathrm{GeV}$, a ``$\ B_c^*$-factory" \cite{Berezhnoy:2016etd}), or hadronic\cite{Baranov:1997wy,Trunin:2015uma,Li:2009ug,Chen:2024dkx}  and photonic\cite{Baranov:1997wy,Dorokhov:2020nvv,Chen:2020dtu}  production mechanisms.
	Finally, beyond QCD corrections for $B_c$-related processes, further calculation of relativistic corrections which have been relatively less studied, will be essential for a deeper understanding of QCD and the properties of $B_c$ mesons.

	\section{Appendix}
	\label{apendx}
 
	In this section, we present the SDC ratios of the relativistic corrections (denoted as G) relative to the LO ones (denoted as F) for the processes $e^+e^-\rightarrow Z^0/\gamma^*\rightarrow (c\bar{b}) [{}^{2S_1+1}{L_1}_{J_1}]+(b\bar{c})[{}^{2S_2+1}{L_2}_{J_2}]$ in the high collision energy limit (  $M^2<<s$),   and we define $R_i=G[n_i]/F[n_1+n_2]$ (where $n_i,i=1,2$ denote the final Fock states) and $\xi=m_c/m_b$ in the following expressions.	
	For the production of double S-wave states, we presented the analytical expressions for different propagation processes, namely $\gamma$-propagator, $Z^0$-propagator and the total of the above two processes including interference effects. 
	
	For ${ }^3S_1^{[1]}+{ }^1S_0^{[1]}$ production, 
	$~Z^0$-propagated and total cross section ratios are the same,
	\bea
	R_1&=& -\frac{11-12\xi+6\xi^2-12\xi^3+11\xi^4}{3(1+\xi)^2(1+\xi^2)}, \\
	R_2&=&-\frac{11-4\xi+6\xi^2-4\xi^3+11\xi^4}{3(1+\xi)^2(1+\xi^2)}.
	\eea
	$\gamma^*$-propagated process cross sections ratios,
	\bea
	R_1&=&-\frac{7-28\xi+3\xi^2-6\xi^3+56\xi^4-14\xi^5}{3(1+\xi)^2(1-2\xi^3)}\\
	R_2&=& -\frac{7-32\xi+3\xi^2-6\xi^3+64\xi^4-14\xi^5}{3(1+\xi)^2(1-2\xi^3)}
	\eea

	For	${ }^3S_1^{[1]}+{ }^3S_1^{[1]}$ production,
	$ Z^0$-propagated process cross section ratios,
	\bea
	R_1&=& R_2 = \frac{
		3 \left(11 - 12 \xi + 6 \xi^2 - 12 \xi^3 + 11 \xi^4\right) - 4 \left(11 - 12 \xi + 9 \xi^2 - 24 \xi^3 + 22 \xi^4\right) \sin^2\theta_w
	}{
		-9 \left(1 + \xi\right)^2 \left(1 + \xi^2\right) + 12 \left(1 + \xi\right)^2 \left(1 + 2 \xi^2\right) \sin^2\theta_w
	}
	\eea	
	$ \gamma^*$-propagated process cross section ratios,
	\bea
	R_1&=& R_2 = \frac{-22\xi^4+24\xi^3-9\xi^2+12\xi-11}{3(\xi+1)^2(2\xi^2+1)}
	\eea
	Total cross section ratios,
 \bea
	R_1 = R_2 &=& \big[ -9 \left(1 + \xi^2\right) \left(11 - 12 \xi + 6 \xi^2 - 12 \xi^3 + 11 \xi^4\right) + 12 \left(1 - \xi^2\right) \left(11 - 12 \xi + 14 \xi^2 - 12 \xi^3 + 11 \xi^4\right) \sin^2\theta_w  \cr
	&&- 8 \left(11 - 12 \xi + 31 \xi^2 - 48 \xi^3 + 67 \xi^4 - 156 \xi^5 + 143 \xi^6\right) \sin^4\theta_w \big]\big/ \big[ 27 \left(1 + \xi\right)^2 \left(1 + \xi^2\right)^2 - 36 \left(1 - \xi^4\right)\cr&& \left(1 + \xi\right)^2 \sin^2\theta_w  + 24 \left(1 + \xi\right)^2 \left(1 + 4 \xi^2 + 13 \xi^4\right) \sin^4\theta_w \big]
	\eea
	
		For	${ }^1S_0^{[1]}+{ }^1S_0^{[1]}$ production,
	$ Z^0$-propagated process cross section ratios,
	\bea
	R_1=R_2= \frac{3(11-4\xi+6\xi^2-4\xi^3+11\xi^4)-4(11-4\xi+9\xi^2-8\xi^3+22\xi^4) \sin^2\theta_w}{-9(1+\xi)^2(1+\xi^2)+12(1+\xi)^2(1+2\xi^2) \sin^2\theta_w}
	\eea
	$ \gamma^*$-propagated process cross section ratios,
	\bea
	R_1=R_2=\frac{-22\xi^4+8\xi^3-9\xi^2+4\xi-11}{3(\xi+1)^2(2\xi^2+1)}
	\eea
	Total cross section ratios,
	\bea 
R_1=R_2&=&[-9(1+\xi^2)(11-4\xi+6\xi^2-4\xi^3+11\xi^4)+12(1-\xi^2)(11-4\xi+14\xi^2-4\xi^3+11\xi^4)\sin^2\theta_w\cr&&-8(11-4\xi+31\xi^2-16\xi^3+67\xi^4-52\xi^5+143\xi^6)\sin^4\theta_w]/[27(1+\xi)^2(1+\xi^2)^2- 36 \left(1 - \xi^4\right)\cr&& \left(1 + \xi\right)^2\sin^2\theta_w+24(1+\xi)^2(1+4\xi^2+13\xi^4)\sin^4\theta_w]
	\eea
 
	One can see when $m_c=m_b$ (i.e. $\xi=1$), these ratios are consistent with that in Refs.~\cite{Li:2013csa,Xu:2012am,Xu:2014zra,Wang:2025sbx}.
	
	For simplicity, we only show the numerical values of ratios $R$ for other processes in Table \ref{Rratios},

\begin{table}
    \caption{The ratios $R$ for the process $e^+e^-\rightarrow  (c\bar{b}) [{}^{2S_1+1}{L_1}_{J_1}]+(b\bar{c})[{}^{2S_2+1}{L_2}_{J_2}]$ in the high collision energy limit ($M^2<<s$). The values are obtained by setting $m_c=1.5 ~\mathrm{GeV},m_b=4.8 ~\mathrm{GeV}$. }
    \begin{tabular}{|c|c|c|c|c|c|c|}
        \hline
        ~   &\multicolumn{2}{c|}{$\gamma^*$-propagated} &  \multicolumn{2}{c|}{$Z^0$-propagated}    &\multicolumn{2}{c|}{ total}  \\
        \hline
        ~&~$R_1$~&$R_2$ & ~$R_1$~&$R_2$& $R_1$&$R_2$ \\
        \hline
        $B_c^{*+}+B_c^{*-}$&$-1.23$ & $-1.23$  & $-1.39$    & $-1.39$ & $-1.28$   & $-1.28$ \\
        \hline
        $B_c^{*+}+B_c^{-}$&$0.24$ & $0.48$  &   $-1.34$  &$-1.82$  &  $-1.34$  &  $-1.82$\\
        \hline
        $B_c^{+}+B_c^{-}$&$-1.72$ &  $-1.72$ &   $-1.87$  & $-1.87$ & $-1.76$   &  $-1.76$\\
        \hline
        $B_c^{*+}+h_{bc}^{-}$& $0.24$ & $-1.15$  &   $-1.61$  &  $-3.22$&   $-1.61$ &  $-3.22$\\
        \hline
        $B_c^{*+}+\chi_{bc0}^{-}$&$-0.21$ &  $-2.47$ & $-0.48$    &$-3.84$  & $-0.23$   & $-2.58$ \\
        \hline
        $B_c^{*+}+\chi_{bc1}^{-}$&$-0.09$ &  $-1.19$ &    $-1.34$ & $-2.47$ &  $-1.34$  & $-2.47$ \\
        \hline
        $B_c^{*+}+\chi_{bc2}^{-}$&$-1.55$ &  $-2.86$ &  $-1.48$   &  $-2.74$&   $-1.53$ & $-2.82$ \\
        \hline
        $B_c^{+}+h_{bc}^{-}$&$-2.28$ &  $-3.45$ &   $-2.02$  & $-3.14$ &$-2.17$    &  $-3.32$\\
        \hline
        $B_c^{+}+\chi_{bc0}^{-}$&  &    &   $-0.82$  & $-3.08$ & $-0.82$   & $-3.08$ \\
        \hline
        $B_c^{+}+\chi_{bc1}^{-}$&$-1.72$  &$-2.37$   &$-1.87$     &$-2.52$  &$-1.76$    &$-2.42$  \\
        \hline
        $B_c^{+}+\chi_{bc2}^{-}$&$1.03$ &  $-1.12$ &   $-1.99$  & $-2.78$ & $-1.99$   & $-2.78$ \\
        \hline	
        \multicolumn{7}{|c|}{t-channel double photon processes} \\
        \hline
        ~   &\multicolumn{3}{c|}{ $R_1$} &	\multicolumn{3}{c|}{ $R_2$}  \\
        \hline
        $B_c^{*+}+B_c^{*-}$	&\multicolumn{3}{c|}{ $-0.89$} &	\multicolumn{3}{c|}{ $-0.89$}  \\
        \hline
        $B_c^{*+}+B_c^{-}$	&\multicolumn{3}{c|}{ $-0.89$} &	\multicolumn{3}{c|}{ $-3.49$}  \\
        \hline
        $B_c^{+}+B_c^{-}$	&\multicolumn{3}{c|}{ $-1.37$} &	\multicolumn{3}{c|}{ $-1.37$}  \\
        \hline	
    \end{tabular}
    \label{Rratios}
\end{table}

	\section{Acknowledgements:} This work was supported by the National Natural Science Foundation of China (No. 11705078, 12575087).

	\hspace{2cm}


\end{document}